\documentclass[aip,reprint]{revtex4-2} 

\usepackage[T2A,T1]{fontenc}
\usepackage[utf8]{inputenc}
\usepackage[russian,english]{babel}
\usepackage{dcolumn}
\usepackage{microtype}
\usepackage{hyperref}
\usepackage{url}
\usepackage{listings}
\usepackage{import}
\usepackage{lipsum} 
\usepackage{siunitx}
\usepackage{gensymb}
\usepackage{graphicx}
\usepackage{subfig}

\usepackage{amsmath}
\usepackage{amssymb}

\usepackage[euler]{textgreek}
\usepackage[version=4]{mhchem}
    \DeclareSIUnit{\molar}{M}
\usepackage{xfrac}

\newcommand*{\fullref}[1]{\hyperref[{#1}]{\autoref*{#1} -- \nameref*{#1}}}

\begin{document}


\title[Journal of Rheology]{Exploring the Nonlinear Rheology of Composite Hydrogels: A New Paradigm for LAOS Analysis} 



\author{Wayan A. Fontaine-Seiler}
\email[waf14@georgetown.edu]{}
\affiliation{Georgetown University, Department of Physics 37\textsuperscript{th} and O Streets NW, Washington, DC 20057, U.S.A.}

\author{Gavin J. Donley}
\email[gavin.donley@nist.gov]{}
\affiliation{National Institute of Standards and Technology (NIST), Gaithersburg, MD 20899, U.S.A.}

\author{Emanuela Del Gado}
\email[edg610@georgetown.edu]{}
\affiliation{Georgetown University, Department of Physics 37\textsuperscript{th} and O Streets NW, Washington, DC 20057, U.S.A.}

\author{Daniel L. Blair}
\email[dlb76@georgetown.edu]{}
\affiliation{Georgetown University, Department of Physics 37\textsuperscript{th} and O Streets NW, Washington, DC 20057, U.S.A.}


\date{\today}

\begin{abstract}
 We investigate composite biopolymer networks composed of co-polymerized fibrin and gelatin networks and perform rheological measurements over a broad parameter space including strain amplitudes that go beyond the linear response regime. One goal of the work presented here is to provide a prototypical biocomposite material with highly separable polymerization times and controlled nonlinear rheological characteristics. We then extend the Sequence of Physical Processes (SPP)\cite{Rogers2017} into a  statistical and geometrical framework that fingerprints the nonlinear rheological response of biopolymer composite gels. Our analysis is based on the changes in the shape of the time-dependent Cole-Cole plots that allow to reduce the time resolved analysis of the SPP method into the transition between linear and nonlinear behavior as the rheological parameters are varied. While the results clearly highlight how the mechanical responses of individual constituents are not simply additive, our extended SPP analysis provides a robust and intuitive classification scheme for comparing the nonlinear response of composite materials subjected to large oscillatory strain.
\end{abstract}


\maketitle 

\section{Introduction}

The demand for new biomaterials with enhanced structural and biochemical functionality for three dimensional bio-printing\cite{gungor-ozkerim_bioinks_2018}, tissue scaffolds and culture environments\cite{rosales_design_2016} requires both the design of multi-component biopolymer networks and the concomitant development of tools to characterize their nonlinear mechanical properties\cite{Buchmann2021}. 
Of particular interest are interpenetrating biologically derived polymer networks that combine inherent physical and biochemical functionality enhanced through the interactions of the individual constituents. In the biological and material science context, most characterization is done at the level of the linear viscoelastic response. However, many processing and printing conditions expose gel networks and embedded cells to large strain deformations and rates that can limit the cellular viability in those\cite{Kim2022,Wang2017} or impact cell behavior\cite{Mierke2011,Kai2016,Parsons2010,Rijns2024}. As a consequence, characterizing strain stiffening or softening, as well as distinctive yielding behavior, of the composites is particularly important. 

In highly specialized applications, the rates of polymerization for each component must be tunable over a broad range of length and time scales in order to optimize functionality. In the case of biopolymer composites, both controlled polymerization kinetics and enhanced nonlinear rheology are important designing aspects to be considered \cite{Webber2007}.
As a matter of fact, the physical or chemical addition of two or more constituents often result in enhanced properties beyond a simple sum of the parts \cite{vereroudakis2020competitive,mugnai2025interspecies}. Hence designing composites with optimized bulk mechanics, modified reactivity, and tunable microscopic architectures requires carefully engineering of how the individual components interact \cite{ROSHANBINFAR2025123174,Rijns2024JACS}, and advanced characterization of their nonlinear rheological response.  
Predicting the emerging nonlinear response of the composite materials is particularly challenging when the role of the microscopic interactions if not fully understood or controlled. Therefore, rheological analytical tools that provide comparative quantitative measures and a more complete characterization beyond the linear response are extremely needed.  



Large Amplitude Oscillatory Shear (LAOS) rheology has emerged as one of the main approaches to 
subjects materials to large shear strains extending Small Amplitude Oscillatory Shear (SAOS) rheology beyond the linear regime, without turning to continuous shear rheological tests where the growth of the imposed strain is unbound, and allowing for a time-resolved exploration of the mechanical response. The challenge is in the quantification and in the interpretation of the effects detected. Therefore, over the past twenty years, there has been a sharp rise in the development of LAOS analysis tools that quantify the nonlinear behaviors seen in these tests. These LAOS analyses can be classified into three main categories differentiated by their approach to handling the nonlinearity of the LAOS waveform: (1) time-based harmonic approaches (Fourier transform rheology \cite{Wilhelm1999,Wilhelm2000,Wilhelm2002}, waveform decomposition \cite{Klein2007}, and Chebyshev polynomial rheology \cite{Ewoldt2008,Ewoldt2009}), (2) geometric strain-vs.-rate decomposition approaches (stress decomposition \cite{Cho2005} and strain decomposition \cite{Dimitriou2013}), and (3) time-derivative trajectory approaches (the sequence of physical processes (SPP) framework \cite{Donley2019,Rogers2012,Rogers2017}).
What remains clear is that the quantitative analysis of LAOS rheology presents a distinct set of challenges that often obscure what can be gained from this powerful rheological measurement method.

In this work, we propose new analytical tools specifically designed to characterize the 
response of a prototypical biopolymer composite gel to  LAOS measurements. The methodology proposed here provides an extension of the SPP framework, which precisely exploits the capacity of LAOS to provide access to the time resolved rheological response of such a complex material. Our approach reduces the parameter space to a set of quantitative rheologically relevant metrics that can be easily compared across systems -- allowing for an intuitive interpretation of LAOS rheology results.

The paper is organized as follows. In Section \ref{background} we provide an overview of the LAOS rheology background that guides our analysis. Section \ref{materials-methods} contains all information on the materials chosen and the measurements performed. In Section \ref{sec:stat_analysis} we present the new analysis tools developed for the time-dependent Cole-Cole plots obtained from the SPP approach for LAOS, and in Section \ref{results} we discuss their application to the LAOS rheology of the biopolymer composite gel of choice. Finally Section \ref{conclusions} contains a summary and concluding remarks. 

\section{Background for LAOS rheology.}
\label{background}

Mechanical models that describe the linear oscillatory response of soft and polymeric materials are so universal that their application to data is often straightforward. 
However, an intuitive understanding of nonlinear rheology is instead very often lacking. 
Large amplitudes oscillations are an extremely facile measurements to probe how materials adapt to strains that explore the nascent nonlinearity, either by probing the transition through different response regimes, or taking the system beyond the point of failure. However, unlike linear response, the types of tools needed to connect the materials response to the micro- to meso-scale changes at these extremes are largely underdeveloped in spite of considerable effort. What follows is a brief review 

When performing non-linear rheology measurements on a rheometer, the fundamental form of the data is the time-series of angular displacement and torque values, which can be converted to shear strain and shear stress, respectively, when the details of the test geometry are well understood. Time derivatives of these data series can additionally provide the rates, accelerations, etc. of these parameters.

By using a periodic test protocol and allowing the response of the material to reach a steady alternance state (as defined as when the periodic response of successive cycles overlap in strain/rate/stress space), we enable the characterization of the resulting response waveforms by decomposition into orthonormal functions. In oscillatory rheology, the use of a periodic sinusoidal input waveform ($\gamma(t)=\gamma_{0}\sin{(\omega t)}$) enables the extraction of viscoelastic parameters such as the dynamic moduli ($G'$ and $G''$), based on the in-phase and out-of-phase portions of the response ($\sigma(t)=G'\gamma(t)+G''\dot{\gamma}(t)/\omega=G'\sin{(\omega t)}+G''\cos{(\omega t)}$). While the analysis of these moduli is relatively well understood when the deformation in the linear regime, where the response function is always some form of phase-shifted sinusoid, the more complicated behaviors seen in LAOS require a generalization of this approach. Existing analyses have used harmonic series to fit the $\sigma(t)$ behavior\cite{Wilhelm1999,Wilhelm2000,Wilhelm2002,Klein2007,Ewoldt2008,Ewoldt2009}, geometric decompositions of $\sigma(\gamma)$ and $\sigma(\dot{\gamma})$\cite{Cho2005,Dimitriou2013}, and time-derivative trajectory approaches in strain-rate stress space $\sigma(\gamma,\dot{\gamma})$\cite{Rogers2011,Rogers2012,Rogers2017}.

The time-based harmonic approaches typically represent the stress as a function of time, using the values of a harmonic series such as a Fourier transform to capture non-linear effects at timescales shorter than the oscillation period \cite{Wilhelm1999,Wilhelm2000,Wilhelm2002}. In this case the dynamic moduli $G'$ and $G''$ are the coefficients of the primary harmonic, with every successive odd harmonic providing additional information about the nonlinearity of the stress behavior. (The even harmonics are typically left out of the rheological analysis, as they are often small and do not follow the 2-fold cyclic symmetry of the shear response at steady alternance. They are typically attributed to experimental noise or secondary effects such as wall slip.) In addition to the use of Fourier transforms, techniques that decompose these harmonics into derived parameters \cite{Klein2007}, or use more specialized series such as Chebyshev polynomials \cite{Ewoldt2008,Ewoldt2009} have also been developed. The primary benefits of the time-based harmonic approaches is their ease of implementation (as a simple series decomposition of a response) and the ability to distinguish subtle mechanical behaviors even when the signal to noise ratio is low \cite{Wilhelm1999}. However, the interpretation of the higher harmonics is quite challenging, as all harmonics are acting in parallel at any given time, and aside from some general behavioral descriptions extracted from the 3rd harmonic \cite{Klein2007,Wilhelm2000}, there has been little progress in connecting the various modes of the response to specific physical features.

As it is 
challenging to provide intuitive and easily communicable interpretations of the harmonic spectrum and coefficients in these approaches, 
it can be difficult to translate this approach to the 
design, composition and conception stages in material engineering where intuitive, fast and relevant feedback from mechanical characterization methods is essential. 
Moreover, series decomposition methods 
convolute the signatures of 
significant transient and time-dependent features 
in the complex rheological behavior by distributing the feature between multiple harmonic modes, particularly if the timescale of the effect is short relative to the length of the period. 

To move beyond the simple harmonic series approaches to something more physically meaningful, the second class of LAOS analysis techniques, namely the stress decomposition of Cho et al.\cite{Cho2005} and strain decomposition of Dimitriou et al.\cite{Dimitriou2013}, utilize geometric approaches on the nonlinear Lissajous curves to analytically decompose the material response into solid-like and fluid-like components. As these techniques do not require all behaviors to be present at all times, they can access some transient features which would be obscured by the harmonic decomposition of the first class of metrics. However, the geometric assumptions behind the the specific decompositions proposed by these techniques have not been explicitly shown to reflect the physics seen in all classes of systems; the strain decomposition for example makes assumptions which do not correctly represent the rheophysics of simple yield stress fluids \cite{Rogers2017}. As such, it is difficult to apply these techniques to complicated systems where the underlying physical behaviors of the material are not well understood.

The final class of LAOS analyses utilize the time-resolved derivatives of strain, rate, and stress to define instantaneous moduli which can evolve as the deformation is applied. 
The primary implementation of this type of time-resolved analysis 
is the Sequence of Physical Processes (SPP) framework of Rogers and collaborators.\cite{Rogers2011,Rogers2017} 
The SPP approach treats the material behavior as a trajectory in strain-rate-stress space, and 
defines transient moduli $G'_t(t)$ and $G''_t(t)$ such that the stress at any given time is:

\begin{equation}
\sigma(t) = G'_t (t) \gamma (t) + \frac{G''_t(t)}{\omega}\dot{\gamma}(t)+\sigma ^{d} (t),
\end{equation}

where the $\sigma ^{d}(t)$ term allows for a vertical offset in stress. The only assumption made by this approach is that the moduli are \textit{instantaneously} linear \cite{Rogers2017}, which holds in most systems but can create apparent fluctuations when the underlying physics is inherently nonlinear (i.e. nonlinear elasticity \cite{Donley2022} or some nonlinear viscous models\cite{Rogers2012a}). In this framework, it can be shown that the values of these moduli are equivalent to the partial derivatives of stress with respect to strain and rate, respectively\cite{Rogers2017,Kamani2023}:

\begin{equation}
G'_t(t)=\frac{\partial\sigma(t)}{\partial\gamma(t)},
\end{equation}
\begin{equation}
G''_t(t)=\omega\frac{\partial\sigma(t)}{\partial\dot{\gamma}(t)}.
\end{equation}

The averages of these transient moduli over a period of oscillation are also analogous to the dynamic storage and loss moduli $G'$ and $G''$, and the transient moduli are also identical to the traditional definitions in the linear regime where mechanical properties remain constant. 

The SPP approach has been applied successfully to interpret 
LAOS rheology across a wide range of materials and models \cite{Armstrong2020,Choi2019,Donley2019,Donley2019_2,Donley2022,Korculanin2021,Lee2017,Rogers2012,Rogers2012a,Rogers2017}, and is the only LAOS analysis framework that has been explicitly shown to correctly interpret the time-resolved rheophysics of soft materials which undergo a yielding-type transition\cite{Donley2019,Kamani2023}.


This analysis framework has also been 
successfully used to identify formulation-relevant features of a material response inaccessible by linear data or time-averaged techniques, in the case of the 
rheological behavior of soldering pastes in an industrial screen printing process\cite{Donley2019_2}. This was possible due to time-resolved comparisons of $G'_t$ and $G''_t$ throughout the imposed cyclic deformation, allowing the temporal location and duration of 
regimes exhibiting 
solid-like 
($G'_t > G''_t$) and fluid-like ($G'_t < G''_t$) behaviors to be distinguished\cite{Donley2019,Donley2019_2}.


Unlike what can be achieved for linear regimes under sinusoidal oscillatory strain, it is impossible to fully reduce arbitrary responses to a finite, let alone restricted, set of physical parameters and variables, without loss of information. Any representation using a limited set of parameters and variables will be therefore necessarily incomplete, partial and unable to fully describe the original data. 

Nonetheless, this does not imply that representations using a restricted set of variables cannot be physically relevant or useful. As with any limited model, representations relying on carefully selected important and physically relevant parameters and variables can help understand and classify the observed behaviors. 

It should 
be noted that, by itself, the SPP framework does not attempt to reduce data to a limited set of parameters but correspond to a transformation of the response curves from a stress/strain/strain rate space to a $(G',G'')$ (Cole-Cole plots) and other described parameter spaces\cite{Choi2019,Donley2019,Rogers2017} with limited loss of information. While some studies have proposed additional time-resolved parameters derived from $G'_t$ and $G''_t$ to tackle specific use cases\cite{Choi2019,Donley2019}, no general set of reduced parameters has been proposed for the SPP framework.

The statistical and geometrical information that can be extracted from the transformation of raw experimental stress/strain/strain rate to Cole-Cole plots ($(G'_t,G''_t)$ plots) gives access to important and quantifiable information about the rheological behavior of materials during LAOS cycle. 
We here propose a statistical and geometrical analysis extension to the SPP framework to build intuition towards the understanding of complex nonlinear behavior under oscillatory strain for novel materials and biomaterials.

\section{Materials \& Methods}
\label{materials-methods}
\subsection{Gel Synthesis}

The complete interpenetrating network was originally designed by \textit{Dubbin et al.} for the 3-dimensional culture of neurons and the study of vascularized systems in-vitro\cite{Dubbin2020}. It is composed of two interpenetrating biopolymer networks: fibrin and gelatin crosslinked by a nonspecific enzymatic crosslinker. These gels are synthesized according to the protocol included with the supplemental materials provided by the original authors of\cite{Dubbin2020}.

The materials used for the synthesis of the complete composite network are the following:
\begin{itemize}
    \item Phosphate Buffer Saline (PBS) 1X (no ions i.e. no calcium or magnesium ions) (Buffer and solvent)
    \item 15\% by weight gelatin solution in PBS 1X No Ions
    \item 50 \si{\milli \gram \milli \liter \tothe{-1}} fibrinogen solution in in PBS 1X No Ions
    \item 250 \si{\milli \molar} \ce{CaCl2} in in PBS 1X No Ions
    \item 60 \si{\milli \gram \milli \liter \tothe{-1}} transglutaminase solution in in PBS 1X No Ions
    \item 200 U/ml thrombin solution in PBS 1X No Ions
\end{itemize}

All the solutions are sterilized via filter sterilization during their preparation and kept at 37\degree C in a beads bath before the beginning of the synthesis of the composite gel.

\subsubsection{Intermediary solutions}

For the needs of the synthesis protocol, the following intermediary solutions are also prepared:
\begin{itemize}
    \item Thrombin solution at 200U\si{\milli \liter \tothe{-1}} in PBS 1X No Ions
    \item \ce{CaCl2} solution at 250\si{\milli \molar} in PBS 1X No Ions
    \item Fibrinogen solution at 50 \si{\milli \gram \milli \liter \tothe{-1}} in PBS 1X No Ions
    \item Transglutaminase solution at  60 \si{\milli \gram \milli \liter \tothe{-1}} in PBS 1X No Ions
    \item Gelatin solution at 15\% by weight
\end{itemize}

\paragraph{Thrombin solution}The thrombin solution is made from lyophilized thrombin from bovine plasma (Sigma-Aldrich T4648-1KU) dissolved and diluted to 200U\si{\milli \liter \tothe{-1}} in PBS 1X No Ions (Gibco Phosphate Buffer Saline 10010-023) and filtered sterilized through 0.22 \si{\micro \meter} PES membrane filter (Millex-GP SLGPR33RS). The resulting solution is store in 1mL aliquots and kept under -20\degree C or below for subsequent use.

\paragraph{Calcium Chloride solution}The \ce{CaCl2} solution is made by disolving Calcium Chloride (\ce{CaCl2}) (Sigma-Aldrich C1016) in PBS 1X No Ions (Gibco Phosphate Buffer Saline 10010-023) and filtered sterilized through 0.22 \si{\micro \meter} PES membrane filter (Millex-GP SLGPR33RS).

\paragraph{Fibrinogen solution} The fibrinogen solution is made by dissolving type 1-S fibrinogen from bovine plasma (Sigma-Aldrich F8630 - 65-85\% protein ($\geq$ 75\% clottable)) in PBS 1X No Ions (Gibco Phosphate Buffer Saline 10010-023). For a final solution of 1ml, 0.05\si{\milli\gram} of fibrinogen powder is added to 9.5\si{\milli \liter} of PBS 1X buffer to account for the volume of the fibrinogen powder in the final solution. The resulting solution is filtered sterilized through 0.22 \si{\micro \meter} PES membrane filter (Millex-GP SLGPR33RS).

\paragraph{Transglutaminase solution} The transglutaminase solution is made by disolving transglutaminase (Modernist Pantry - Moo Gloo TI transglutaminase formula) in PBS 1X No Ions (Gibco Phosphate Buffer Saline 10010-023). For a final solution of 1ml, 0.06\si{\milli\gram} of fibrinogen powder is added to 9.4\si{\milli \liter} of PBS 1X buffer to account for the volume of the transglutaminase powder in the final solution. The resulting solution is filtered sterilized through 0.22 \si{\micro \meter} PES membrane filter (Millex-GP SLGPR33RS).

\paragraph{Gelatin solution}The synthesis of the gelatin solution at 15\% by weight is outlined in \cite{Afewerki2018}. The synthesis protocol is summarized below:
\begin{enumerate}
    \item 75\si{\milli \liter} of PBS 1X No ions is added to a sterile bottle ($\geq 150\si{\milli \liter}$).
    \item 15\si{\milli \gram} of type A gelatin from porcine skin (Sigma-Aldrich G1890) are added to the bottle.
    \item The solution is placed at 60\degree C and stirred for 2 hours.
    \item After all the power is dissolved, the pH is adjusted using a 1N solution of sodium hydroxide (\ce{NaOH}) (Fisher Chemical SS277 or Sigma-Aldrich S2770) to reach a precise value between 7.50 and 7.54 (not physiological pH).
    \item The necessary quantity of PBS 1X No ions is added to reach a solution volume of 100\si{\milli \liter}
    \item The solution is filtered sterilized through a vacuum filter setup with a 0.22\si{\micro \meter} PES filter element (VWR 28199-792 or equivalent) and stored at 4\degree C for later use.
\end{enumerate}

\subsubsection{Composite gel synthesis}

The complete composite gel is composed of the following materials and solutions prepared in advance and synthesized according to this protocol and the quantities specified below are for a total final volume of 1.0 ml. The composite gel composition was scaled as needed depending on the volume needed for our measurements. All the solutions are warmed and kept at 37\degree C in a beads bath prior and during their use at the exception of the thrombin solution which remains stored at $\leq -20$\degree C.
\begin{enumerate}
    \item 200 \si{\micro \liter} of fibrinogen solution at 50 \si{\milli \gram \milli \liter \tothe{-1}} is added to a 15\si{\milli \liter} conical tube.
    \item 257 \si{\micro \liter} PBS [-] Ions are added to the 15\si{\milli \liter} conical tube and gently mixed by pipette.
    \item 33.3 \si{\micro \liter} the TG solution at 60 \si{\milli \gram \milli \liter \tothe{-1}} are added to the 15\si{\milli \liter} conical tube and gently mixed by pipette.
    \item 10 \si{\micro \liter} of \ce{CaCl2} solution at 250 \si{\milli \molar} are added to the 15\si{\milli \liter} conical tube.
    \item The solution in the 15\si{\milli \liter} conical tube is left to rewarm for 5 minutes.
    \item 500 \si{\micro \liter} of the gelatin solution is added to the 15\si{\milli \liter} conical tube and thoroughly mixed by means of a Pasteur pipette.
    \item The solution in the 15\si{\milli \liter} tube is left at 37\degree C for 15 minutes.
    \item During the 15 minutes wait, the thrombin solution is warmed up (by hand) and 5\si{\micro \liter} are placed in a 50\si{\milli \liter} conical tube.
    \item At the end of the 15 minute wait, the solution from the 15\si{\milli \liter} conical tube is added to the 50\si{\milli \liter} tube. The resulting solution is mixed carefully and thoroughly by pipette over a few seconds.
    \item The solution is immediately loaded in the rheometer and the plate geometry lowered.
\end{enumerate}

\subsubsection{Partial gels synthesis}
\label{gel_syn}
Partial gels were synthesized according to the equivalent protocols, total volumes, procedures and waiting times between individual steps. We also replace the missing constituent volumes with buffer solution (PBS 1X [-]ions). This method of synthesizing the partial gels was chosen to maintain identical concentrations for the remaining constituents of the gels between all of gels and partial gels studied. The compositions considered for our measurements and analysis are summarized in Table \ref{table:composition}.

\begin{table}[ht!]
        \begin{tabular}{ccc}
        Composition& Transglutaminase & No Transglutaminase\\
        \hline \hline  
        Gelatin & G+ & --\\
        Fibrin& F+& F\\
        Gelatin-Fibrin& GF+& GF\\
        \end{tabular}
        \caption{Partial gels synthesis summary. The shorthand for each partial gel will follow the values in the table above. Transglutaminase + denotes the addition of transglutaminase to the gel. Note that gelatin without transglutaminase is not considered in this work. \label{table:composition} }
\end{table}

The case of gelatin alone without transglutaminase (G) is considered outside of the scope of this work due to its mechanical properties at our experimental conditions. At a temperature of $37\degree$C, gelatin at a 7.5\% mass concentration does not form a gel and remains a viscous liquid.

\subsection{Rheological Methods}

An Anton-Paar MCR-702 stress-controlled rheometer, with parallel plate-plate geometry in dual motor configuration is used for all the rheological measurements. In this configuration, the bottom plate applies the strain and the top plate measures the stress. The bottom plate is 50mm in diameter and the top plate is a parallel plate with a diameter of 25mm. The plate-plate geometry is encased in a clamshell temperature controlled hood with air circulation to maintain a sample at an ambient temperature of 37\degree C during the rheological measurements. All measurements are carried out with a gap of 1\si{\milli \meter} between the top and bottom plate of rheometer.

The samples are sealed on their edges, 10 minutes after being loaded in the rheometer, with light mineral oil (Fisher Chemicals 0121-1) to prevent water evaporation and dehydration of the gels during the measurements. This time of 10 minutes was chosen to ensure the samples did not mix with the mineral oil applied to their edge before they have time to polymerize sufficiently as to prevent contamination of the samples with mineral oil. The only exceptions to this procedure were the partial gels which did not contain thrombin (i.e. GF \& G gels) due to their slower initial polymerization rate. These partial gels were sealed at after 90 minutes to allow for additional polymerization time.

\subsubsection{Small Amplitude Oscillatory Strain}
\label{sec:SAOS_theor}

The rheological characterization of the synthesized composite gel and partial gels is carried out using a Small Amplitude Oscillatory Strain (SAOS) measurement by subjecting the gels to sinusoidal oscillatory strains $\gamma(\gamma_0, \omega,t) = \gamma_0 \sin(\omega t )$.

Strain sweeps were performed on each sample with a fixed angular frequency of $\omega = 2\pi$ rad/s
for the applied oscillatory strain $\gamma(\gamma_0, \omega) = \gamma_0 \sin(\omega t + \varphi)$. Frequency sweeps were also performed on each sample with a fixed strain amplitude of $\gamma_0 = 1\%$ for the applied oscillatory strain.

\subsubsection{Polymerization kinetics}

We monitor the evolution of $G'$ and $G''$ of the individual gel and partial gels over time after loading using SAOS at strain amplitude of $\gamma_0 = 10^{-2} = 1\%$ and an angular frequency of $\omega = 2\pi$ rad/s 
Samples were considered fully polymerized after 3 hours.

\subsubsection{Continuous shear}

Continuous shear measurements were carried out on fully polymerized samples by applying a constant shear rate $\dot{\gamma}=10^{-4}$ 1/s 
with the resulting stress measured every second until the cumulative strain $\gamma=\int_0^T \dot{\gamma} \mathrm{d}t$ reached a value of $10^2$.

\subsubsection{Large Amplitude Oscillatory Strain}

Large Amplitude Oscillatory Strain measurements were carried out on each individual gel and partial gel after full polymerization. For each combination of angular frequency and strain amplitude described below, the average of both stress/strain and stress/strain rate response curves after steady oscillatory responses were recorded for later analysis.

The LAOS "Frequency Sweeps" consisted of individual recordings of the Lissajoux curves as described above with fixed strain amplitudes of for individual strain amplitude values of $0.2$ and $0.4$ and for individual angular frequencies ranging from $\frac{2\pi}{20}$ and $40\pi2$ rad/s 

The LAOS "Strain Sweeps" consisted of individual recordings of the Lissajoux curves for fixed angular frequencies of $\frac{2\pi}{5}$ and $\frac{2\pi}{2}$ rad/s 
and for individual strain amplitude values increasing from $10^{-2}$ to $2\cdot 10^2$ for both angular frequencies.

\begin{figure*}[htpb]
    \includegraphics[width=0.90\linewidth]{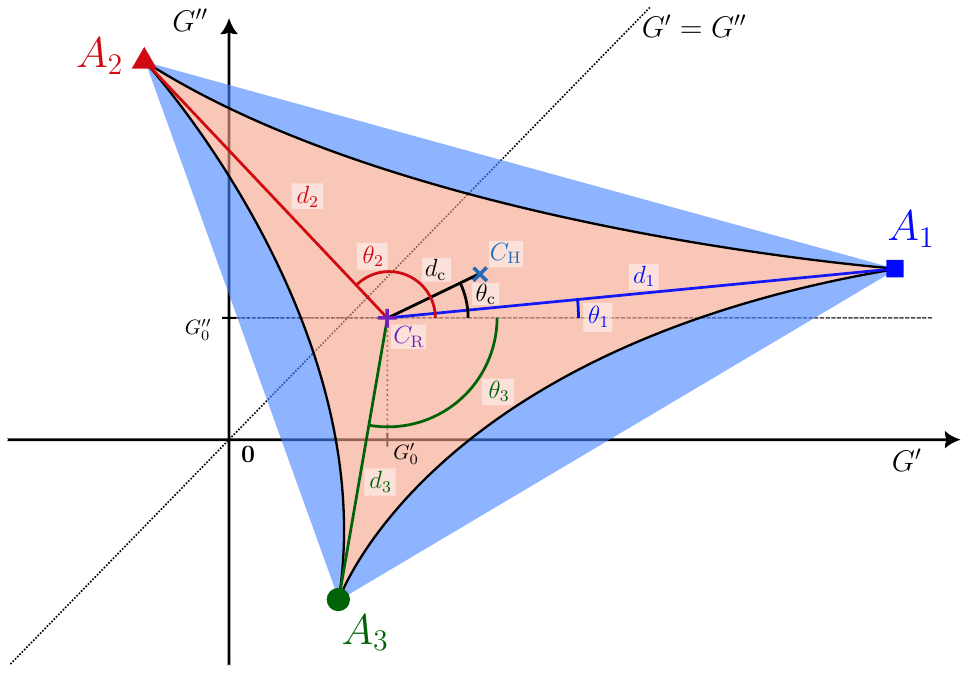}
    \caption{\label{fig:spp_legend} Legend and description of the geometrical measurements carried out on the temporal Cole-Cole parametric plots. The apices, $A_1$, $A_2$ and $A_3$, are defined as the most extreme points of the deltoid formed by the parametric plot, i.e. the points forming the convex hull for the individual points of the deltoid. The rheological center ($C_{\mathrm{R}}$) correspond to the average point of all the points of the deltoid. The values of $(G_0',G_0'')$ at this point are also equal to the storage and loss moduli derived from the fundamental Fourier frequency of the stress-strain response and from SAOS rheology. The hull center ($C_{\mathrm{H}}$) correspond to the barycenter of the apices forming the convex hull of the deltoid. The area of the deltoid correspond to the area in salmon and to the numerical integration of the surface delimited by the points of the deltoid. The hull area, represented by the addition of the salmon and blue areas, correspond to the area formed by the apices triangle. 
    The polar coordinates of the apices and hull center with 
    respect to the rheological center are also shown in the color of their respective points. The diagonal dotted line corresponds to the points for which the transient moduli 
    are equal (i.e. $G'_t=G''_t$).}
\end{figure*}

\section{LAOS-SPP Analysis Framework}
\label{sec:stat_analysis}
In previous work, the analysis of the parametric Cole-Cole plots has primarily focused on time-resolving the specific physical events which occur within a period of oscillation\cite{Donley2019_2,Lee2017}. When derived parameters have been defined, they have typically also been time-resolved parameters 
such as a time resolved-phase angle ($\delta _t = \tan^{-1}(G''_t/G'_t)$)\cite{Donley2019}.
While the fundamental origins of the deltoid geometry typically seen in the Cole-Cole plots has previously been explored\cite{Choi2019}, 
a more comprehensive statistical characterization of 
the geometry of these 
parametric plots remains undeveloped. We hypothesize a general form of the Cole-Cole plots for a given material. In what follows, we will describe our new framework through the analysis of the parametric plots over a range of applied of strain amplitudes at a fixed frequency for each composition described in Section \ref{gel_syn}. Using the analysis outlined 
below, we 
identify easily interpretable parameters that elucidate the transition from linear to nonlinear response as the strain amplitude is systematically increased.

\label{subsubsec:legend}

For the purpose of analysis, we consider the following spaces:
\begin{itemize}
    \item The strain, rate and stress 3-dimensional response space $\mathcal{U} = \mathbb{R}^3$,  set of all values $[\gamma , \dot{\gamma}, \sigma] \in \mathbb{R}^3$, with $\overrightarrow{U}$ its associated vector space
    \item The closed moduli space $\mathcal{G} = \mathbb{R}^2$, set of all values $[G',G''] \in \mathbb{R}^2$, with $\overrightarrow{G}$ its associated vector space. $\left[ 
\mathbf{\hat{g}'},\mathbf{\hat{g}''} \right]^{T}$ an orthonormal basis of the $\mathcal{G}$ space and $\hat{\mathbf{g'}}$ and $\hat{\mathbf{g''}}$ unit vectors along the $G'$ and $G''$ directions respectively
    \item The time space $\mathcal{T} = [0,T], T\in \mathbb{R}_+$, the period of the sinusoidal strain applied
\end{itemize}

The SPP processing of the strain response can be represented by a function/transform $\mathcal{Y}$ from $\mathcal{U} * \mathcal{T}$ to $\mathcal{G} * \mathcal{T}$. The general resulting parametric plot from the application of $\mathcal{Y}$ on the stress response $\sigma(t)$ to the strain $\gamma(\gamma_0, \omega,t) = \gamma_0 \sin(\omega t + \varphi)$ forms a closed curve $\mathcal{C}_G$ in $\mathcal{G}$, set of all $\mathbf{s}=\mathbf{G}_{\mathrm{tr}} = \left[ \mathbf{G}_{\mathrm{tr}} \cdot \hat{\mathbf{g'}} ,\mathbf{G}_{\mathrm{tr}} \cdot \hat{\mathbf{g''}} \right]^{T}\in \mathcal{G}$ over $\mathcal{T}$. The closed curve $\mathcal{C}_G$ also defines a surface $\mathcal{S}_G$ in $\mathcal{G}$ 

By considering the time dependence of the parametric curves over $\mathcal{T}$, we can also write that the closed curve $\mathcal{C}_G$ is parametrized in time by $\mathcal{P}_G(\mathbf{s})$ the corresponding time to each $\mathbf{s} \in \mathcal{C}_G$. $\mathcal{C}_G$ is formed by the temporal points over a period $T$, $\mathbf{G}_{\mathrm{tr}} (t)$ with $t\in \mathcal{T} = \left[0,T \right]$ with $D_t(\mathbf{s}) =  \left( \frac{\mathrm{d}\mathcal{P}_G}{\mathrm{d}\mathbf{s}} \right)^{-1}$ the temporal density over the curve $\mathcal{C}_G$ such that $\oint_{\mathcal{C}_\mathrm{G}} \mathrm{d}\mathbf{s} D_t(\mathbf{s}) = \int_0 ^{T} \mathrm{d}t \mathbf{G}_{\mathrm{tr}} (t)$.

Since experimental data are composed of discrete points, we can also define the discrete case with the set of all transient moduli values $\mathbf{G}_{\mathrm{tr}}(n)$, linearly sampled points in time with $n\in[1,N]=\mathcal{N} $ the set of discrete sequential experimental sampling points linearly spaced in time such that $t(n) = n\cdot \frac{T}{N}$.

From these discrete points, we can directly compute several geometrical and statistical quantities to extend the application of the SPP framework. To characterize the geometrical and statistical aspects of the parametric plots, we calculate the parameters outlined below to provide a set of rheological measures that can be compared across materials and relative concentrations of constituents. We will define each parameter presented in Figure \ref{fig:spp_legend} and provide mathematical derivation in what follows. 

An exhaustive list of the statistical and geometrical quantities computed is presented below:
\begin{itemize}
    \item $\mathbf{C}_{\mathrm{R}}$ -- Rheological Center 
    \item $\mathbf{C}_{\mathrm{H}}$ -- Hull Center
    \item $\mathbf{A}_p$ with $p \in [1,3]$ -- Apices of the deltoid shapes
    \item $[S_\mathrm{curve}, \tilde{S}_{\mathrm{curve}}]$ -- Deltoid area and normalized area 
    \item $[S_{\mathrm{hull}},\tilde{S}_{\mathrm{hull}}]$ -- Hull area and normalized area
    \item $\mathcal{R}$ -- Ratio of deltoid and hull areas $\sfrac{S_\mathrm{curve}}{S_\mathrm{hull}}$
    \item $\left[ d_\mathrm{c}, \theta_\mathrm{c} \right]$ -- the polar coordinate of the vector $\mathbf{\underline{C}}$ between $\mathbf{C}_{\mathrm{R}}$ and $\mathbf{C}_{\mathrm{H}}$
    \item $\left[ d_p, \theta_p \right]$ with $p \in [1,3]$ -- the polar coordinates of the vectors $\mathbf{\underline{A}_p}$ from $\mathbf{C}_{\mathrm{R}}$ to each apex $\mathbf{A}_{p}$
\end{itemize}

For the purpose of illustration, a typical specimen plot is presented in figure \ref{fig:spp_legend} with the statistical and geometrical quantities described above represented.

\subparagraph{Rheological center}
The rheological center $\mathbf{C}_{\mathrm{R}}$ corresponds to the time weighted average of the closed curve $\mathcal{C}_\mathrm{G}$. In the continuous case, we define it as:
\begin{eqnarray}
    \mathbf{C}_{\mathrm{R}} = \frac{1}{L} \oint_{\mathcal{C}_\mathrm{G}} \mathrm{d}\mathbf{s} \cdot \mathbf{\hat{n}_s} \ \mathbf{s} \ D_t(\mathbf{s}) 
\end{eqnarray}
with $L=\oint_{\mathcal{C}_\mathrm{G}} \mathrm{d}\mathbf{s}\cdot \mathbf{\hat{n}_s} \ D_t(\mathbf{s})$, with $\mathbf{\hat{n}_s} = \frac{\mathbf{s}}{|\mathbf{s}|}$ the time-weighted contour length of the close curve $\mathcal{C}_\mathrm{G}$.

By considering the time parametrization of $\mathcal{C}_G$, we have the following equivalent statements for $\mathbf{C}_{\mathrm{R}}$:
\begin{eqnarray}
    \mathbf{C}_{\mathrm{R}} = \frac{1}{T} \int_0 ^{T} \mathrm{d}t \ \mathbf{G}_{\mathrm{tr}} (t)
\end{eqnarray}

In the case of experimental data, we approximate $\mathbf{C}_{\mathrm{R}}$ by the discrete sum:
\begin{eqnarray}
    \mathbf{C}_{\mathrm{R}} \simeq \frac{1}{N} \sum_{n=1} ^{N} \mathbf{G}_{\mathrm{tr}} (n)
\end{eqnarray}
corresponding to the unweighted average of $\mathbf{G}_{tr}$ overall all sampling points.

 $\mathbf{C}_{\mathrm{R}}$ is equivalent to the values of storage and loss moduli, $G'(\gamma, \omega)$ and $G''(\gamma,\omega)$ that would be obtained for fixed applied frequency and amplitude when decomposing the fundamental harmonic of the oscillatory stress response over SAOS storage and loss moduli $G'$ and $G''$. Consequently, as will be detailed in future sections, this point serves as a natural reference for vectors points of origin and for the normalization of physical parameters and other statistical quantities we derived from the time resolved moduli response.

\subparagraph{Apices} The apices $\mathbf{A}_p$ with $p \in \mathbb{N}$ of $\mathcal{C}_G$ correspond to the vertices of the convex hull of $\mathcal{C}_G$ in $\mathcal{G}$. These points therefore represent the outermost extension of the parametric plot and therefore the temporal response in $\mathcal{G}$ and the most extreme behaviors a material exhibits during a strain cycle.

The values of the storage and loss moduli of the apices $\mathbf{A}_p$ can be normalized by the storage and loss moduli values respectively of the rheological center such that the normalized apices $\mathbf{\tilde{A}}_p = \left[ \sfrac{\mathbf{A}_p \cdot \mathbf{\hat{g'}}}{\mathbf{C}_\mathrm{R} \cdot \mathbf{\hat{g'}}},\sfrac{\mathbf{A}_p \cdot \mathbf{\hat{g'}}}{\mathbf{C}_\mathrm{R} \cdot \mathbf{\hat{g''}}} \right]^\mathrm{T}$.

Their intrinsic evolution for a given material, as well as their concomitant/combined use with the rheological center ($\mathbf{C}_{\mathrm{R}}$) in order to calculate vector parameters defining distances and orientation between [apices and centers], will allow us to observe, measure quantify both the the linear to nonlinear transition as well as characteristic behavior.

\subparagraph{Hull center}
The hull center $\mathbf{C}_{\mathrm{H}}$ is defined as the discrete average of the apices $\mathbf{A}_p$:
\begin{equation}
\mathbf{C}_{\mathrm{H}} = \frac{1}{P} \sum_{n=1} ^{P} \mathbf{A}_p
\end{equation}
with $P$ the total number of apices

The hull center thus corresponds to the barycenter of the apices $\mathbf{A}_p$ as define previously. Physically, it thus represents the mean of the extreme strain behaviors in the cycle.

\subparagraph{Deltoid Area} 
We consider the area $\mathcal{S}_G$ formed by the closed curve $\mathcal{C}_G$ For $\mathbf{A}\in \mathcal{S}_\mathrm{G}$, we define $S_{\mathrm{curve}}$ the total area defined by the closed line $\mathcal{C}_G$: 
\begin{eqnarray}
    S_{\mathrm{curve}} = \oint _{\mathcal{S}_{G}} \mathrm{d}\mathbf{A}
\end{eqnarray}
Consequently, integration over all $\mathbf{s} \in \mathcal{C}_G$ yields the following expressions for $\mathcal{S}_G$
over $\mathcal{C}_G$:
\begin{eqnarray}
    S_{\mathrm{curve}} & = & \frac{1}{2} \oint _{\mathcal{C}_G} (\mathrm{d}\mathbf{\hat{g}''} \cdot \mathbf{\hat{g}''} ) \cdot (\mathbf{s}\cdot \mathbf{\hat{g}'})\nonumber \\ 
    & = & -\frac{1}{2} \oint _{\mathcal{C}_G} (\mathrm{d}\mathbf{\hat{g}'} \cdot \mathbf{\hat{g}'} ) \cdot (\mathbf{s}\cdot \mathbf{\hat{g}''}\nonumber \\
    & = & \frac{1}{4} \oint_{\mathcal{C}_G} ((\mathrm{d}\mathbf{\hat{g}''} \cdot \mathbf{\hat{g}''} ) \cdot (\mathrm{d}\mathbf{\hat{g}'} \cdot \mathbf{\hat{g}'} ) - (\mathrm{d}\mathbf{s}\cdot \mathbf{\hat{g}'}) (\mathbf{s} \cdot \mathbf{\hat{g}''})) \nonumber \\
    & = & \frac{1}{2} \int _{0} ^{T} \mathrm{d}t \ \left( \mathbf{G}_\mathrm{tr}(t)\cdot \mathbf{\hat{g}''} \right) \left( \frac{\mathrm{d}}{\mathrm{d}t}\left( \mathbf{G}_\mathrm{tr}(t)\cdot \mathbf{\hat{g}'} \right) \right)
\end{eqnarray}
can be obtained from Green-Ostrogradski and divergence theorem. The $\sfrac{1}{2}$ factor is required due to the symmetry of the stress response curve. As the $\mathbf{G}_{\mathrm{tr}}(n)$  will take half a period $T$ to complete the closed curve $\mathcal{C}_G$, the path is completed twice over a full period of the oscillatory stress response, thus yielding a value of $2 S_{\mathrm{curve}}$ when integrating the path of a full period.

In the discrete case, we can numerically approximate the value of the surface $\mathcal{S}_G$ through a discrete trapezoidal integration of the experimental points obtained from the time-resolved analysis of the experimental oscillatory stress response. In this case, by letting $n$ be defined$\mod{N-1}$, we can compute the following approximation for the value of the area of $S_{\mathrm{curve}}$:
\begin{eqnarray}
S_{\mathrm{curve}} & \simeq & \frac{1}{2} \sum_{n=0} ^{N-1} \left[ \frac{1}{2} \left[ \left(\mathbf{G}_{\mathrm{tr}}(n+1)+\mathbf{G}_{\mathrm{tr}}(n) \right) \cdot \mathbf{\hat{g}''}\right] \right. \nonumber \\
 & & \cdot \left[ \left(\mathbf{G}_{\mathrm{tr}}(n+1)-\mathbf{G}_{\mathrm{tr}}(n) \right) \cdot \mathbf{\hat{g}'} \right] \bigg]
\end{eqnarray}

The area $S_{\mathrm{curve}}$ can be normalized by the magnitude of the squared modulus of the rheological center $\mathbf{C}_{\mathrm{R}}$ squared: $\left| \mathbf{C}_{\mathrm{R}} \right|^2 = \left( \mathbf{C}_{\mathrm{R}} \cdot \mathbf{\hat{g}'} \right)^2 + \left( \mathbf{C}_{\mathrm{R}} \cdot \mathbf{\hat{g}''} \right)^2$:
\begin{equation}
    \tilde{S}_{\mathrm{curve}} = \frac{S_{\mathrm{curve}}}{\left| \mathbf{C}_{\mathrm{R}} \right|^2}.
\end{equation}

This normalization factor allows us to factor out run-to-run variations within concentrations, while also allowing for straightforward comparisons between materials of differing composition and storage and loss moduli, thus providing a convenient way to collapse data on single master curve.

The normalized area $\tilde{S}_{\mathrm{curve}}$, provides a quantitative way to determining opening and widening of the area inscribed be the discrete parametric plot. The increasing normalized area can be easily interpreted as a increase of the nonlinear part of the stress response exhibited by the sample\cite{Donley2019_2} and provides a simple quantification method to facilitate comparisons between different frequencies, strain amplitudes or composition.

\subparagraph{Hull Area} The hull area $S_{\mathrm{hull}}$ corresponds to the area of the convex hull of $\mathcal{C}_G$.

As with the area defined by $S_{\mathrm{curve}}$, the hull area can be normalized by the squared modulus of the rheological center $\left| \mathbf{C}_{\mathrm{R}} \right|^2$:
\begin{equation}
    \tilde{S}_{\mathrm{hull}} = \frac{S_{\mathrm{hull}}}{\left| \mathbf{C}_{\mathrm{R}} \right|^2}
\end{equation}

As for the parametric plot area, the hull area and its normalized value can be used as an alternative method of quantifying the nonlinearity of a given sample, albeit one focused more on the extremes of variation. It can also serve to quantify the 2-dimensional footprint area of a given $\left[ G'_t (n), G''_t (n) \right]$ parametric plot considering only the its extrema points.


\subparagraph{Area ratio} Additionally, the ratio between the two areas described above (the deltoid area and 
hull area) can be calculated for a measure of how nonlinear a response is compared to its extrema. This ratio is defined as:
\begin{equation}
    \mathcal{R} = \frac{S_{\mathrm{curve}}}{S_\mathrm{hull}}
\end{equation}

The area ratio highlights and quantifies of the concavity of a parametric plot in respect to its convex hull and their deviation from the simple triangle forming its convex hull. It contributes to the determination of the signature behavior of the material being measured.

\subparagraph{Polar coordinates of the center vectors} 
Considering the points $\mathbf{C}_{\mathrm{R}}$ and the $\mathbf{C}_{\mathrm{H}}$, we can compute the vector from the rheological center to the hull center ($\mathbf{\underline{C}} = \overrightarrow{\mathbf{C}_{\mathrm{R}}\mathbf{C}_{\mathrm{H}}}$) in $\mathcal{G}$:
\begin{equation}
    \mathbf{\underline{C}} = \mathbf{C}_{\mathrm{H}} - \mathbf{C}_{\mathrm{R}}
\end{equation}

The polar coordinates of $\mathbf{\underline{C}}$ can later be computed into the norm and radial components $\left[ d_\mathrm{c}, \theta_\mathrm{c} \right]$ of the vector. The distance $d_\mathrm{c}$ can be further normalized by the norm of the modulus of the rheological center $ \left| \mathbf{C}_{\mathrm{R}} \right| = \left[ \left( \mathbf{C}_{\mathrm{R}} \cdot \mathbf{\hat{g}'} \right)^2 + \left( \mathbf{C}_{\mathrm{R}} \cdot \mathbf{\hat{g}''} \right)^2 \right] ^{\frac{1}{2}}$ to obtain the normalized distance $\tilde{d_\mathrm{c}}$ such that:
\begin{equation}
    \tilde{d_\mathrm{c}} = \frac{d_\mathrm{c}}{\left| \mathbf{C}_{\mathrm{R}} \right|}
\end{equation}

Our experimental data shows that most of the time is spent during a cycle is spent in the vicinity of the extrema regions, i.e., near the individual apices. Comparatively, the transitions between these regions account for a minority of the time it takes to complete a cycle.

Considering these observations, the magnitude and direction of the vector between the rheological center and the hull center can provide approximate insight in the time imbalance between rheological regions our sample exhibit or the deviation from equal time being spent in each region of the $\mathcal{G}$.

In the case where the magnitude of the vector is negligible compared to our normalization factor, approximately equal time is spent between the different regions. 

A norm of the vector significant compared to our normalization factor is the result of an imbalance in time spent between the regions and, therefore, a behavior dominating over the entire cycle and the value of the norm, a quantification of this imbalance. The direction of the vector will provide the direction of the region where most time is spent, and therefore, a qualification of the dominating behavior. This, in turn can help identify the dominant behavior and transitions between different dominant behaviors over a given parameter variation.

The parameters of this vector can provide insights into the nonlinearity as well as the symmetry of the parametric plot of the transient moduli in the $\mathcal{G}$ space.

\subparagraph{Polar coordinates connecting rheological center to apices}
A similar analysis as to the one done for the vector between the rheological center to the hull center is performed on the vectors defined from the rheological center to each individual apex. We therefore define the vectors from the rheological center to each apex $\mathbf{\underline{A}}_p = \overrightarrow{\mathbf{C}_{\mathrm{R}}\mathbf{A}_{p}}$ in $\mathcal{G}$:
\begin{equation}
    \mathbf{\underline{A}}_p = \mathbf{C}_{\mathrm{H}} - \mathbf{A}_{p}
\end{equation}

The coordinates of the apices $\mathbf{\mathbf{A}}_p$ are later converted in polar coordinates in the norm / radial component of the vector $\left[ d_p, \theta_p \right]$. The distance $d_p$ can be further normalized by the norm of the complex modulus of the rheological center $\left| \mathbf{C}_{\mathrm{R}} \right| = \left[ \left( \mathbf{G}_0 \cdot \mathbf{\hat{g}'} \right)^2 + \left( \mathbf{G}_0 \cdot \mathbf{\hat{g}''} \right)^2 \right] ^{\frac{1}{2}}$ to obtain the normalized distance $\tilde{d_p}$ such that:

\begin{equation}
    \tilde{d_p} = \frac{d_p}{\left| \mathbf{C}_{\mathrm{R}} \right|}
\end{equation}

The coordinates of these vectors provide insights on the mechanical characteristics of the most extreme instantaneous responses of the material during a strain cycle. They form the set of the distance and orientation of the ``limit'' instantaneous behaviors exhibited by the sample compared in respect to the average mechanical response in the form of the rheological center $\mathbf{C}_{\mathrm{R}}$. Additionally, when the relative locations of the apices have a high degree of symmetry, they serve as a marker of the potential for instantaneously nonlinear behaviors\cite{Rogers2012a,Donley2022}. 

\section{Results}
\label{results}

\subsection{Small Amplitude Oscillatory Rheology}

To characterize the mechanical behavior the composite hydrogel, we use Small Amplitude Oscillatory Rheology (SAOS) with an applied oscillatory strain, $\gamma(\gamma_0, \omega) = \gamma_0 \sin(\omega t)$. The stress response is decomposed into the in-phase (storage) and out-of-phase (loss) moduli, $G'(\gamma_0, \omega)$ and $G''(\gamma_0, \omega)$, respectively. To quantify how the shear moduli respond to variations in applied strain, we fix the oscillation frequency $(\omega = 6.2$ rad/s 
over a range of strain amplitudes $(\gamma_0 = 0.01 - 25\%)$.  The response of the shear moduli to variations to the applied frequency are found by fixing the strain amplitude, ($\gamma_0 = 1.0\%$) and varying the applied oscillation frequency over the range considered $(\omega = 0.6 - 200$ rad/s).

\begin{figure*}[htpb]
  \centering
  \subfloat{\includegraphics[width=0.50\textwidth]{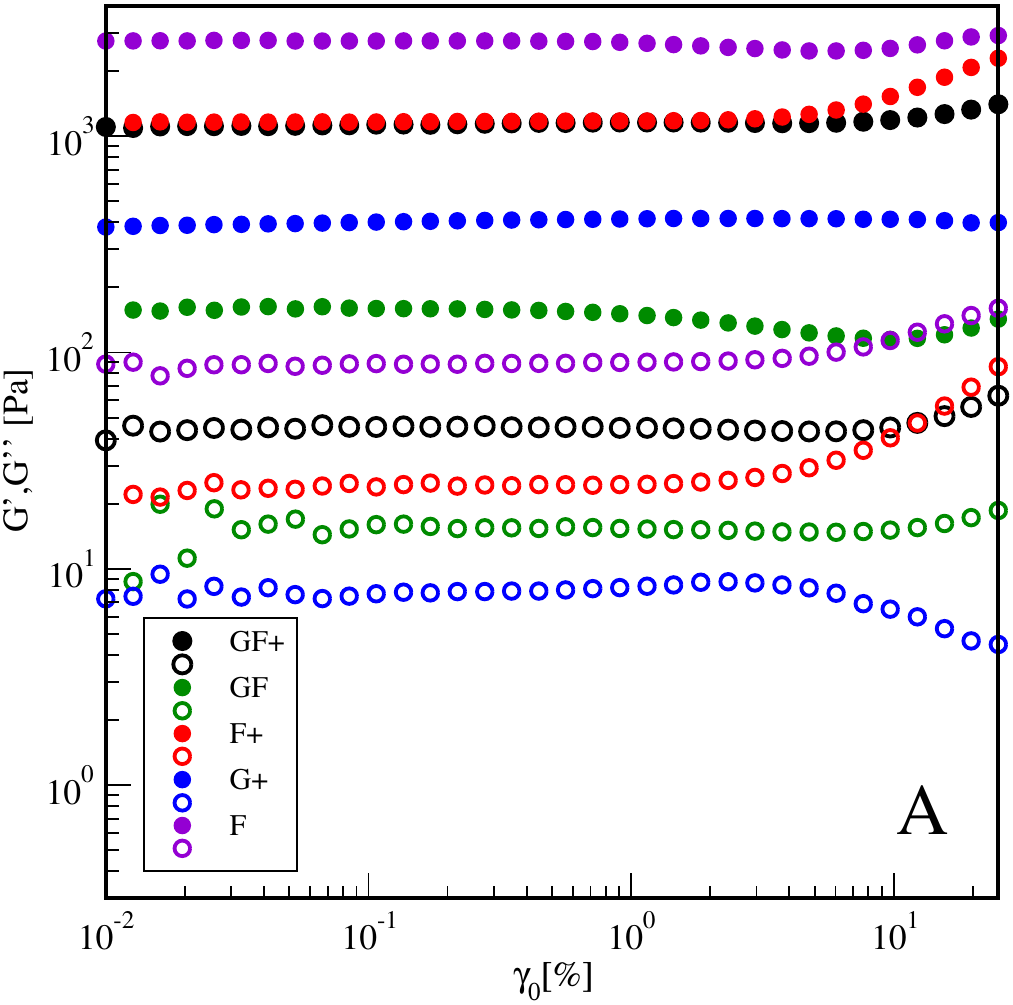}\label{subfig:strain_sweep}}
  \hfill
  \subfloat{\includegraphics[width=0.473\textwidth]{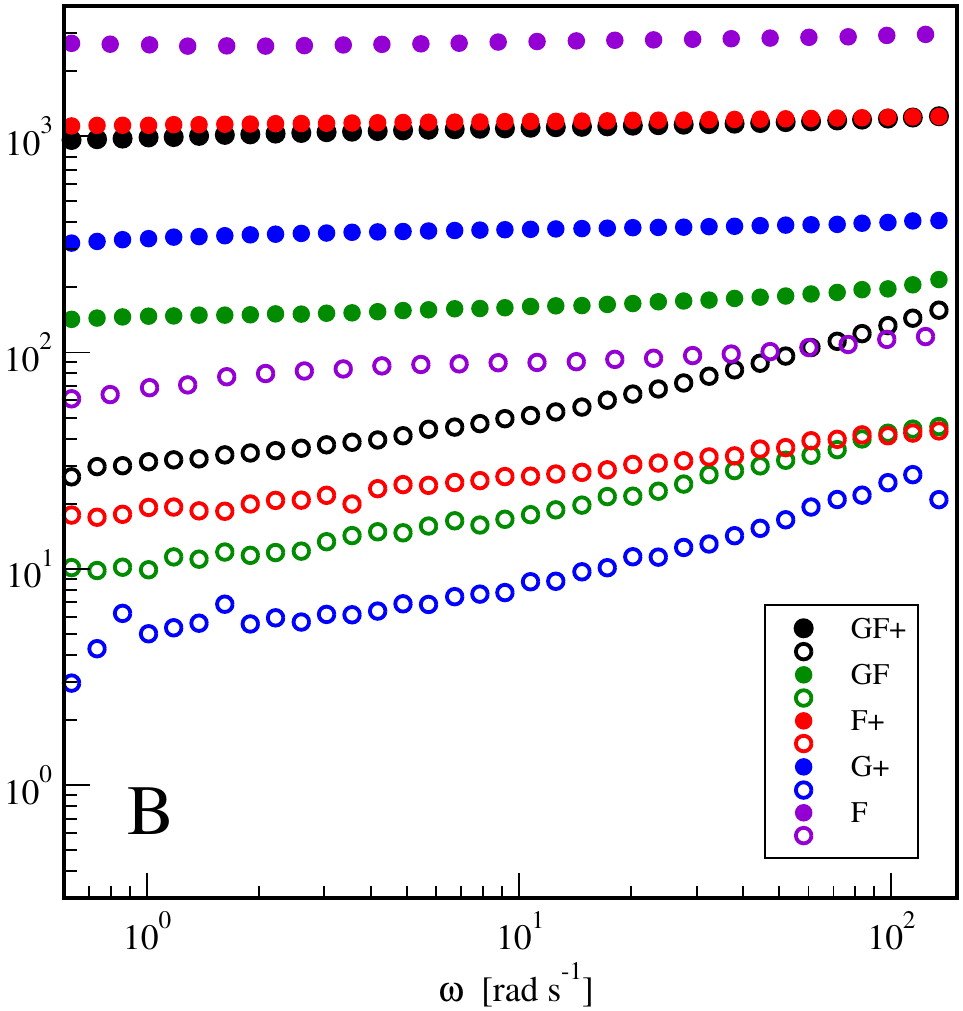}\label{subfig:freq_sweep}}
  \caption{Small amplitude oscillatory strain and frequency sweeps for the GF+, GF, F+, G+ and F compositions as described in table \ref{table:composition}. Closed vs. open symbols indicate the value of $G'(\gamma_0,\omega)$ vs. $G''(\gamma_0,\omega)$, respectively. \textbf{(A)} The strain sweep is performed at a $\omega = 6.2$ rad/s 
  and exhibits a largely linear response over several orders of magnitude of applied strain amplitude $\gamma_0$. \textbf{(B)} Frequency sweep for each gel composition at $\gamma_0 = 1.0\%$ for both the strain amplitude $\gamma _0$ (A) and the frequency for all different formulations. Over the both strain amplitudes and frequency ranges, the oscillatory response of the gel remains dominated by the storage modulus with loss factors $\tan(\delta)$ over $10^2$.}
  \label{figure:lin_rheo}
\end{figure*}

The mechanical response of the composite gel and its constituent components (with and without crosslinking), to variations in $\gamma_0$ at fixed $\omega$ are shown in Figure \ref{figure:lin_rheo}A. These data exhibit a broad range of nearly constant moduli with magnitudes $G' \sim 10^2$ to $10^3 \,[\rm{Pa}]$ and $G'' \sim 10^1$ to $10^2 \,[\rm{Pa}]$ over the entire range of applied strain amplitudes, resulting in a $\tan(\delta) = G''/G' \sim 10^1$ indicating that all of the gels exhibit predominantly elastic behavior for the full range of applied strains. Moreover, each gel only exhibits minor variations in $G'$ at the highest applied strain values, indicating that the maximum strain applied is far below the yield strain for all gels. Additionally, we quantify the variation of the various gel moduli as a function of frequency with a fixed $\gamma_0 = 0.1\%$ and find that all gels exhibit a weak power-law increase ($G' \sim \omega^{0.1}$) over two orders of magnitude in frequency while also showing a large separation between $G'$ and $G''$ for all $\omega$ (Figure \ref{figure:lin_rheo}B). A nearly frequency independent elasticity is typical for single component branched and crosslinked biopolymer networks as seen in work by \textit{Gardel et al.}\cite{Gardel2004}.

Though illustrative of the range of linear elastic behavior, SAOS alone cannot provide insights into the role that reorganization of internal microstructure can play in these interpenetrating and crosslinked gel networks. To further elucidate the connections between microstructure and bulk material properties, large strain amplitudes and nonlinear rheology is required.

\subsection{Continuous Shear}

\begin{figure}[ht!]
    \centering
    \includegraphics[width=1.00\linewidth]{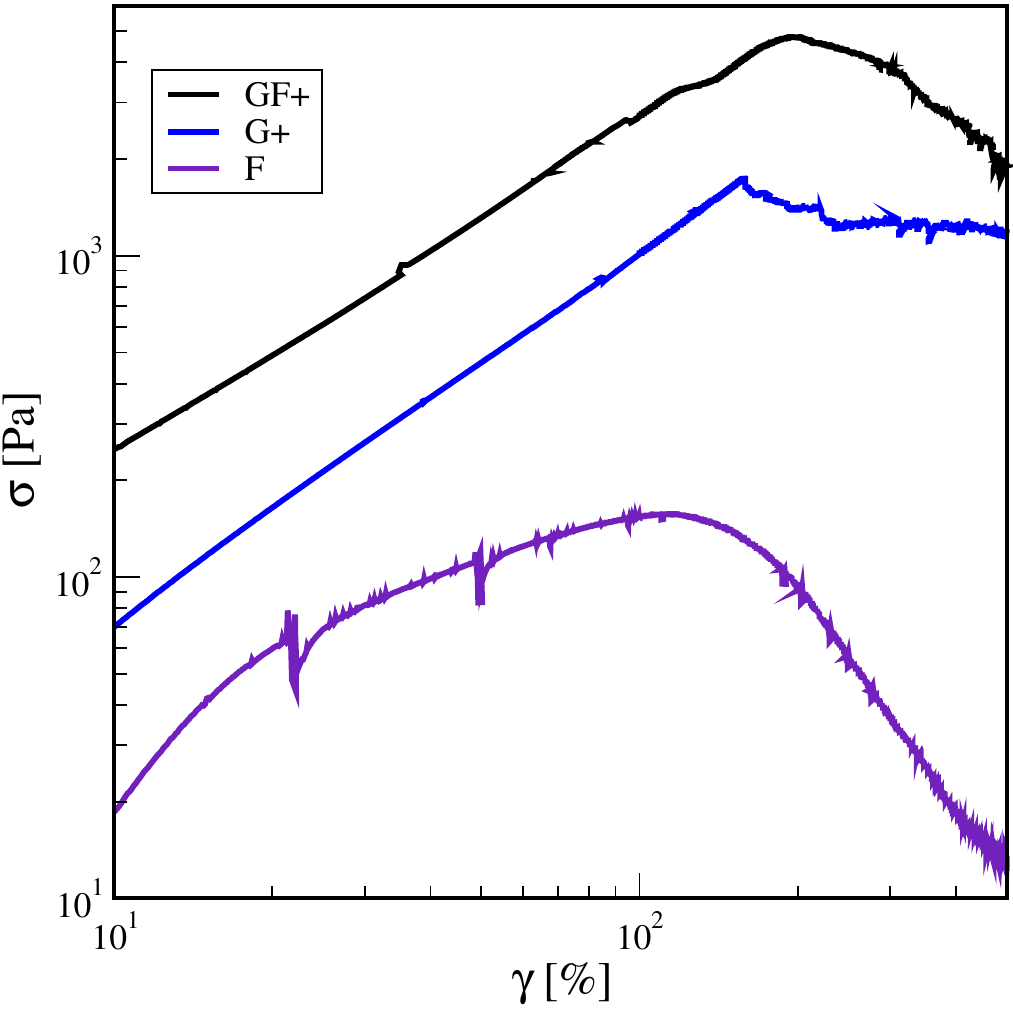}
    \caption{Stress-strain response under constant shear at $\dot{\gamma} = 0.1$ \%/s 
    for the complete gel, the gelatin network and the fibrin network. We can observe, in the case of the full composite gel, the high yielding strain of around 200\% and the difference of behavior between the individual networks and the interpenetrating networks. The fibrin network shows an early but long softening and yielding regime where the gelatin network exhibits a linear elastic response and brittle rupture at strains above 100\%. The composite gel shows a higher yielding strain that both network taken individually.}
    \label{fig:steady_state_shear}
\end{figure}

Continuous shear deformations can reveal the transition from linear to nonlinear behavior\cite{Bischoff2004}, and at large applied strains, the mechanism of failure\cite{Janmey2008}. Crosslinked and branched biopolymer gels like actin and collagen exhibit moderate nonlinear stiffening at relatively modest applied strains $\gamma \sim 10\%$, whereas fibrin gels are often observed to show a large secondary elastic response prior to yielding due primarily to stretching within the network \cite{Kang2009}. To investigate the how the addition of TG impacts the stress-strain response, we apply a continuously increasing strain to a maximum value of $\gamma_0=$1000\% while measuring the shear stress. The response curves presented in Figure \ref{fig:steady_state_shear} follow behaviors that are typical of elastic solids. Interestingly, we do not observe a substantial transition to nonlinear strain stiffening, but instead find that the composite gel and gelatin appear to have linear moduli up their yield strain. Gelatin yields in a way that is consistent with brittle fracture above $\gamma \sim 100\%$. The fibrin response is less linear over the range of applied strains and indicates a reduction in the modulus at moderate strains, and perhaps a stretching of the network at large strain values.

To fully characterize the composite and constituent gels, measurements and analysis must go beyond SAOS and continuous shear experiments. Pure fibrin gels and composites \cite{janmey_jor_83,shah_strain_1997} exhibit nonlinear rheology that is distinctive compared to other biopolymer systems. Large yield strains due in part to hierarchical structural motifs\cite{piechocka_structural_2010} distinguish fibrin gels from many other biopolymers. Therefore, methods that reveal the mechanics at large strains must be utilized. The composite and constituents all exhibit large linear elasticity $\tan(\delta) \sim 10^{-2}$ indicating that Large Amplitude Oscillatory Rheology (LAOS) will provide a  to compare these materials. 

\subsection{Large Amplitude Oscillatory Rheology} 

\begin{figure}[htpb]
  \centering
  \subfloat{\includegraphics[width=0.49\textwidth]{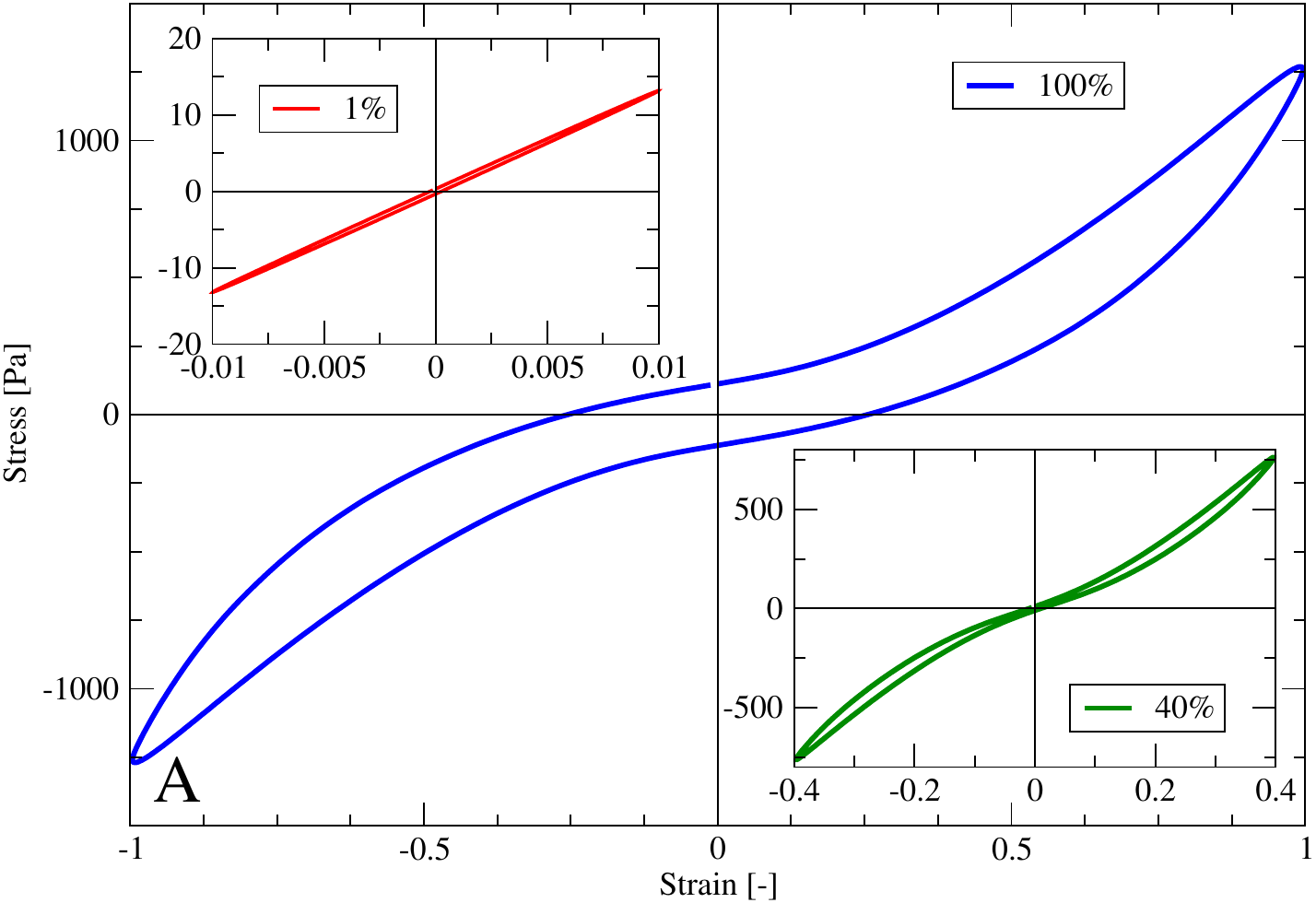}\label{subfig:waveforms}}
  \hfill
  \subfloat{\includegraphics[width=0.49\textwidth]{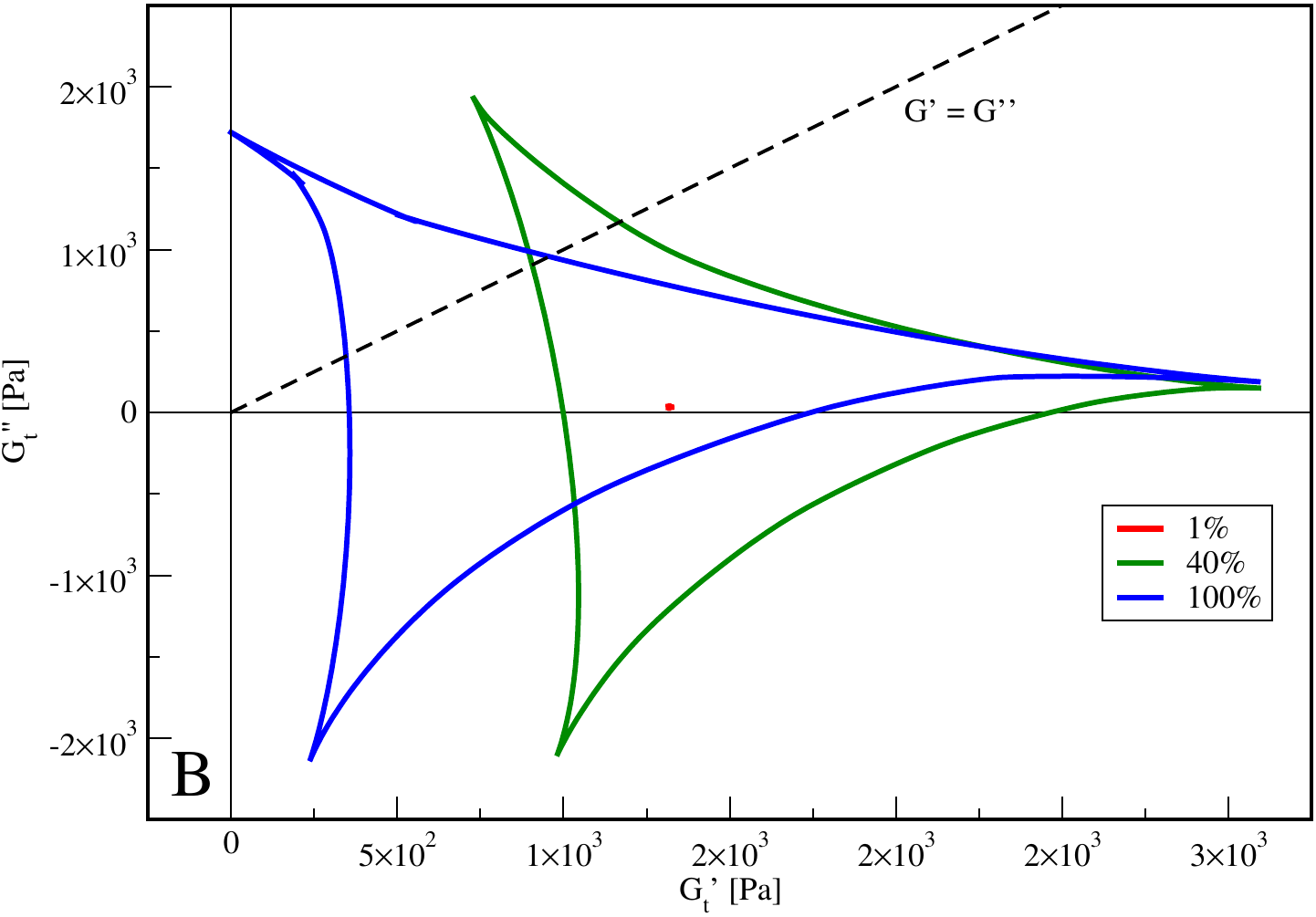}\label{subfig:spp_example}}
  \caption{Experimental stress-strain waveforms/responses for the complete composite biogel after complete polymerization at a frequency of 0.2\si{\hertz} and strain amplitudes $\gamma _0$ of 1\%, 40\% \& 100\%. We can observe the onset and increase in nonlinear behavior as the strain amplitude $\gamma _0$ increases. At a strain amplitude of 1\%, our material exhibits a nearly linear elastic behavior. As the strain amplitude increases, the stress strain responses show an increase in hysteresis and opening of the response curve.}
  \label{fig:waveforms}
\end{figure}

We perform Large Amplitude Oscillatory Rheology (LAOS) on the composite network and its constituents to explore the rheological behavior of beyond the linear regime. Figure \ref{fig:waveforms} shows the raw Lissajous-Bowditch waveforms (stress vs. strain) for the (GF+) gel after polymerization. The frequency  is fixed at $\omega = 1.25$ rad/s
for the strain amplitudes of $\gamma _0$ of 1.0\%, 40\% \& 100\%. To obtain the physical parameters $G'_t$ and $G''_t$ from the experimental oscillatory strain responses, the raw strain responses from the Anton-Paar MC702 were processed using Matlab code provided by \textit{Donley et al.}\cite{Donley2019,Donley2019_2,Donley2021} and the statistical analysis outlined in section \ref{sec:stat_analysis} utilize in-house tools developed in Python. Incorporated in these analysis tools are Fourier domain filtering methods that reduce the impact of experimental noise on the final results. Fourier domain filtering removes the higher harmonics while retaining the more significant lower harmonics allowing us to determine the effective strain response. Additionally, the even harmonics are removed from the response so that the filtered response maintains the two-fold cyclic symmetry of the strain input. The cut-off for the Fourier harmonic filtering is determined separately for each strain response by selecting harmonics of intensity higher than $\sim 10^{-3}I_1$ with $I_1$ the intensity of the fundamental harmonic frequency applied or higher than $10^1$ times the noise floor, whichever is the highest. This enables us to maximize the retention of information from the raw experimental strain responses while retaining exploitable signal to noise ratio filtered responses for the computation of the transient moduli $G'_t$ and $G''_t$. 

\label{subsec:SPP_deltoids}

\begin{figure}[htb]
    \centering
    \includegraphics[width=1.00\linewidth]{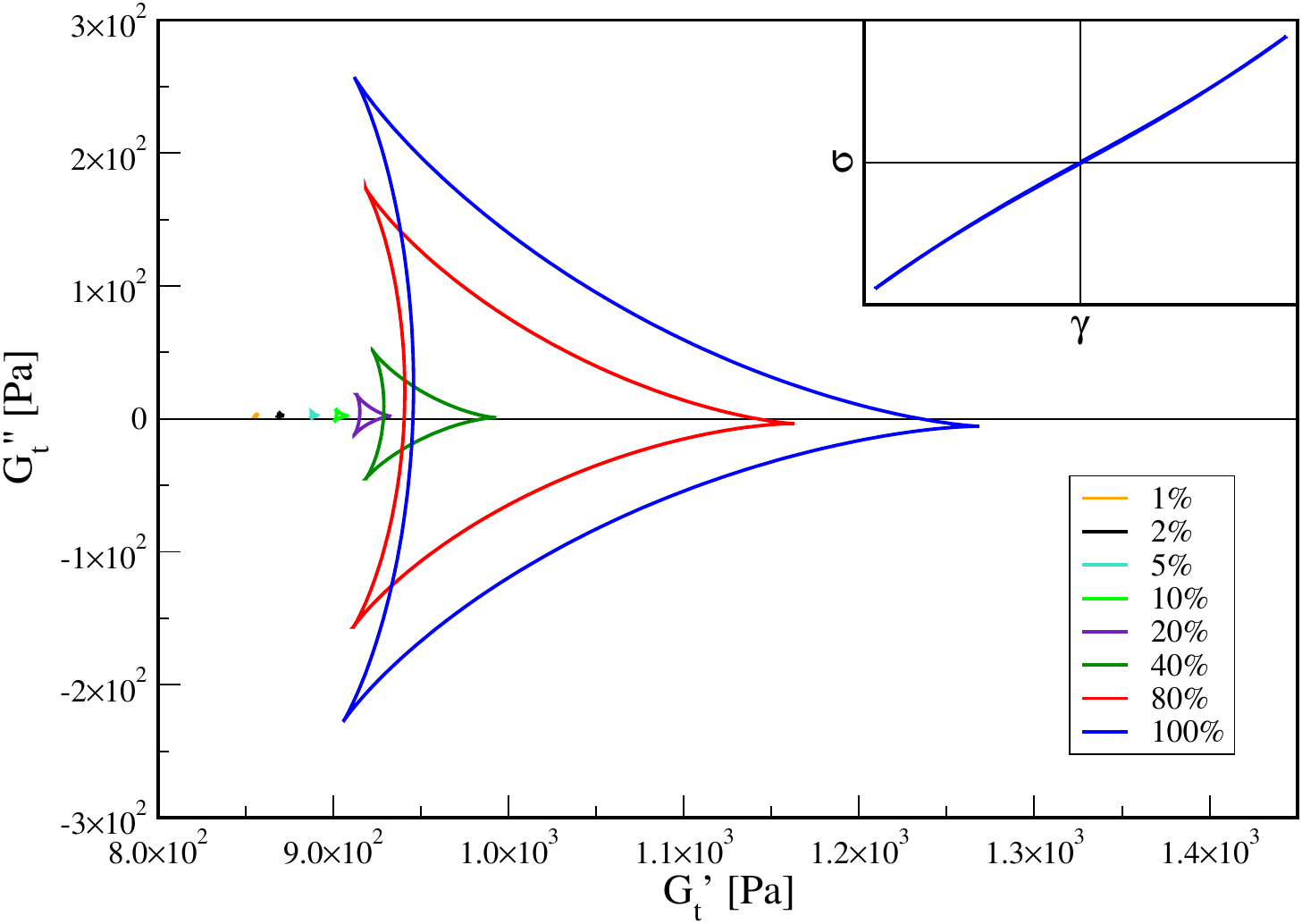}
    \caption{Cole-Cole plot of $G_t'$ and $G_t''$ for stress strain rate response curves at 0.2 \si{\hertz} for strain amplitudes ranging from 1\% to 100\% for the gelatin sub-network. In the top right hand corner, the shape of the stress strain response curve is plotted for a strain amplitude of 100\%. We notice the difference of order of magnitude between $G'_t$ and $G''_t$ with $G'_t\gg G'_t$. This demonstrates that the gelatin network remains almost purely elastic, albeit nonlinear, an observation that is further confirmed by the stress strain response.}
    \label{fig:SPP_gelatin}
\end{figure}
The LAOS response of the composite gel, and the SPP analysis shown in Figure \ref{fig:waveforms} exhibit behavior that we will analyze in the sections to follow. While new insights are gained from the statistical and geometrical tools, distinguishing the contribution (or lack thereof) of each constituent gel in the overall rheological response of the composite is essential to proposing a microscopic physical model of the nonlinear rheology of the composite gel. What's to follow is a complete LAOS-SPP analysis based on our geometric interpretation of the Cole-Cole plots that result from the SPP framework for Gelatin-TG (G+), Fibrin (F) and the Full Gel (GF+).

\paragraph{Gelatin: G+}

In this section we describe the LAOS rheology and SPP analysis for gelatin crosslinked with transglutaminase (G+). The SPP analysis is presented in Figure \ref{fig:SPP_gelatin} along with a representative Lissajous curve for $\gamma_0= 100\%$  (Fig.\- \ref{fig:SPP_gelatin} inset). The high degree of elasticity is reflected in the shape, location and evolution of the parametric plots as a function of applied strain as demonstrated by two observations; the ratio between the extrema of the $G''_t$ values during a cycle, and that the magnitudes of the ranges of $G''_t$ never exceed $\sim 250 \si{\pascal}$ whereas the averages of $G'_t \simeq 10^3 \si{\pascal}$. A low $\frac{\left| G'_t \right|}{G''_t}$ ratio indicates that the material exhibits a highly elastic response which is reinforced by the fact that all of the data exist very far from the $G'_t = G''_t$ line.

\begin{figure*}[htpb]
    \centering
    \includegraphics[width=1.00\linewidth]{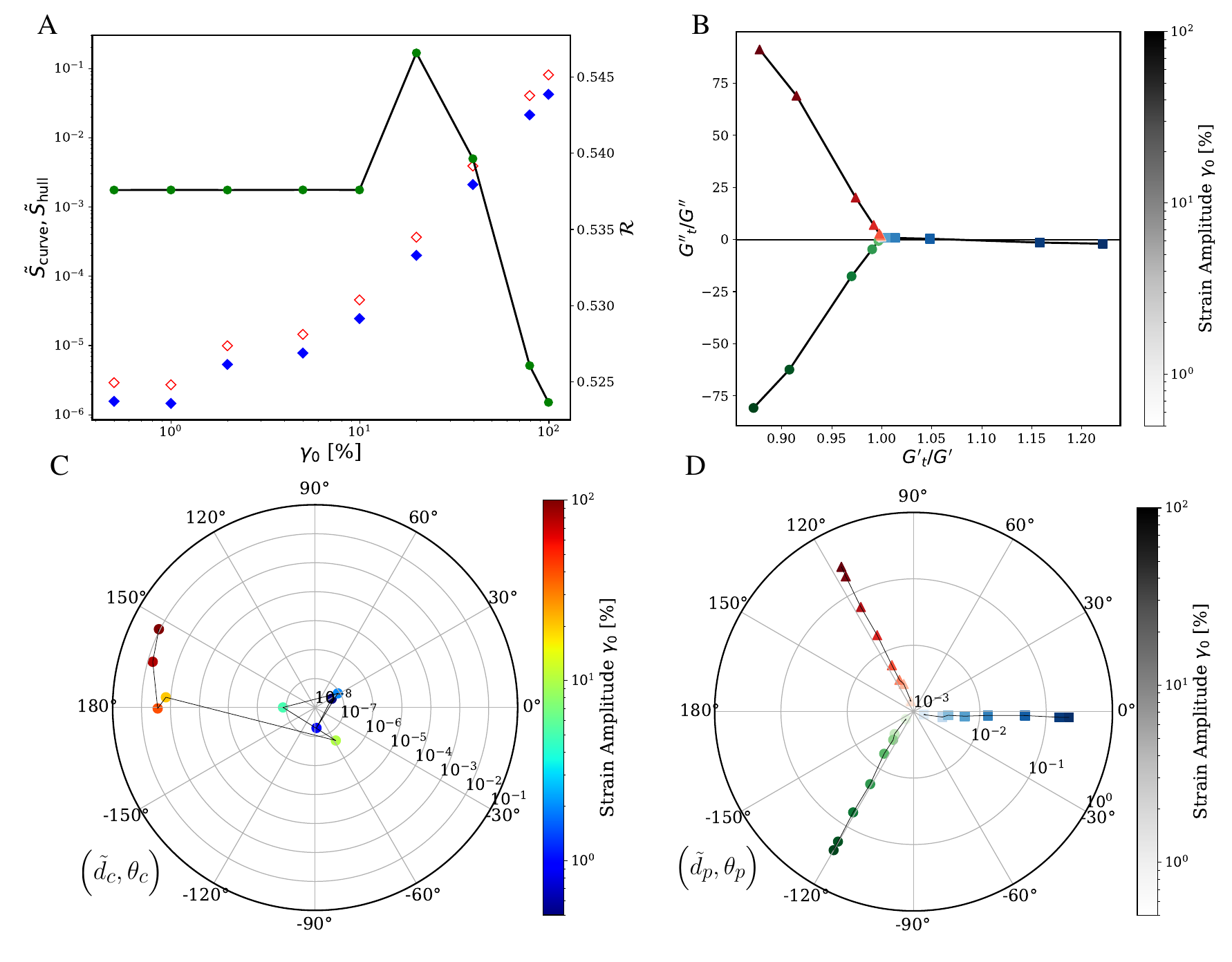}
    \caption{Evolution of the normalized measurements and values described in fig.\ref{fig:spp_legend} for the G+ partial gel (Gelatin only) as a function of strain amplitude. \textbf{(A)} Normalized areas of the parametric response, the convex hull (open red symbols) and parametric plot area (blue symbols) to hull area ratio (green symbols) as defined in fig.\ref{fig:spp_legend}. As the strain amplitude increases, we notice the gradual increase of both the the parametric plot and hull areas but an overall decrease of their ratio with strain amplitude. \textbf{(B)} Normalized storage and loss moduli of the apices $\mathbf{\tilde{A}}_p$ as a function of the strain amplitude of the oscillatory shear imposed to the sample. The evolution of the position of the apices with strain amplitude can be observed and leads to an increasing normalized distance between each apex and the increasing extension of the parametric plot. \textbf{(C)} Polar plot of the coordinates $\left[ \tilde{d}_c , \theta _c \right]$ of the vector between the rheological center ($\mathbf{C}_{\mathrm{R}}$) and the hull center ($\mathbf{C}_\mathrm{H}$) as defined in fig.\ref{fig:spp_legend}. We can observe the increasing distance between the rheological centers and hull centers as the strain amplitude increases. We can also observe that the angle of the vector remains nearly constant at around $\sim180\degree$ such that the vector extends in one direction with the strain amplitude. \textbf{(D)} Plots of the coordinates $\left[ \tilde{d}_p , \theta _p \right]$ of the apices vectors $\mathbf{\underline{A}}_p$ from the rheological center to each individual apex as described in fig.\ref{fig:spp_legend} for the complete composite gel. We can observe the increasing distance between the rheological center and the apices as the strain amplitude increases and the the evolution of the angles. The angles are distributed at angles $0\degree$, $120\degree$ and $-120\degree$ and these values remain nearly constant across the range of strain amplitude applied to our sample.}
    \label{fig:spp_analyzed_gelatin_unified}
\end{figure*}

Our geometrical interpretation of the SPP analysis for (G+) gels (Fig.\- \ref{fig:SPP_gelatin}) will provide the characteristic response of a system that is predominantly elastic over a broad range of applied strains. In Figure \ref{fig:spp_analyzed_gelatin_unified}A the normalized areas of the SPP deltoids $(\tilde{S}_{\rm curve})$, the convex hull constructed from the apices $(\tilde{S}_{\rm hull})$, and their ratio $(\cal R)$ are plotted as a function of strain. We observe an increase of the normalized areas, ranging from $10^{-6}$ at $\gamma_0 = 0.5\%$ to $10^{-1}$ at $\gamma _0 = 100\%$. The ratio $\cal R$ remains constant from $\gamma _0 = 0.5\% \to 20\%$ and exhibits a decreasing trend after $\gamma _0 = 40\%$, indicating an increasing concavity of the paths between apices. The relative expansion and symmetry of the deltoids is quantified by normalizing the magnitude of the time dependent moduli at each apex by the value of the linear moduli (see Fig.\- \ref{fig:spp_analyzed_gelatin_unified}B) as described in Section \ref{subsubsec:legend}. It is important to note that the magnitude of the increase in the normalized loss modulus is not due to an increase in a viscous response, but is directly attributed to a increased instantaneous nonlinear elasticity arising from finite extensibility 
of the material at large strain, as signaled by the near-perfect symmetry of the plots across the $G'_t$ axis \cite{Donley2022}.

The evolution of the SPP-deltoid position and shape encodes the deviation from linear viscoelasticity and the relative magnitude of the elastic to viscous responses, respectively. The polar coordinates $\left[ d_\mathrm{c}, \theta_\mathrm{c} \right]$ of the vector $\mathbf{\underline{C}}$ described in Section \ref{subsubsec:legend} are plotted in Figure \ref{fig:spp_analyzed_gelatin_unified}C  for applied strains ranging from $\gamma _0 = 0.5\to 100\%$. There are two distinct features that become clear from the analysis. First, for strains below $10\%$, the magnitude $d_c$ is $\sim6$ decades smaller than the maximum, implying a primarily linear elastic response over this range of strains. Second, $\theta_c$ maintains an angle very close to $180\degree$, indicating minimal rotation of the time-weighted deltoid away from the $G_t$-axis. Additionally, the polar coordinates $[d_p,\theta_c]$ of the vectors between $R_c$ and each apex $\mathbf{A}_p$, shown in Figure \ref{fig:spp_analyzed_gelatin_unified}D show a highly symmetric orientation that increases in step with the magnitude of the applied strain. Taken together, this behavior is indicative of a purely elastic gel with minimal transition to viscoelasticity even at the highest strain values with only minor strain stiffening occurring at high strains\cite{Donley2022}.

\paragraph{Fibrin: F}

In the conditions we have studied, the polymerization kinetics of gels formed from fibrinogen and thrombin are two orders of magnitude faster than gelatin, while the average mesh size of the final fibrin gel is 3-4 orders of magnitude greater than gelatin. This separation of time and length scales is what makes the systems preferable for constructing a composite. We would anticipate that these microstructural differences should result in distinct rheological characteristics that differentiate fibrin from gelatin over all applied strain values. Using our geometrical analysis of the SPP data, we directly compare the LAOS signatures for each network.

\begin{figure}[htb]
    \centering
    \includegraphics[width=1.00\linewidth]{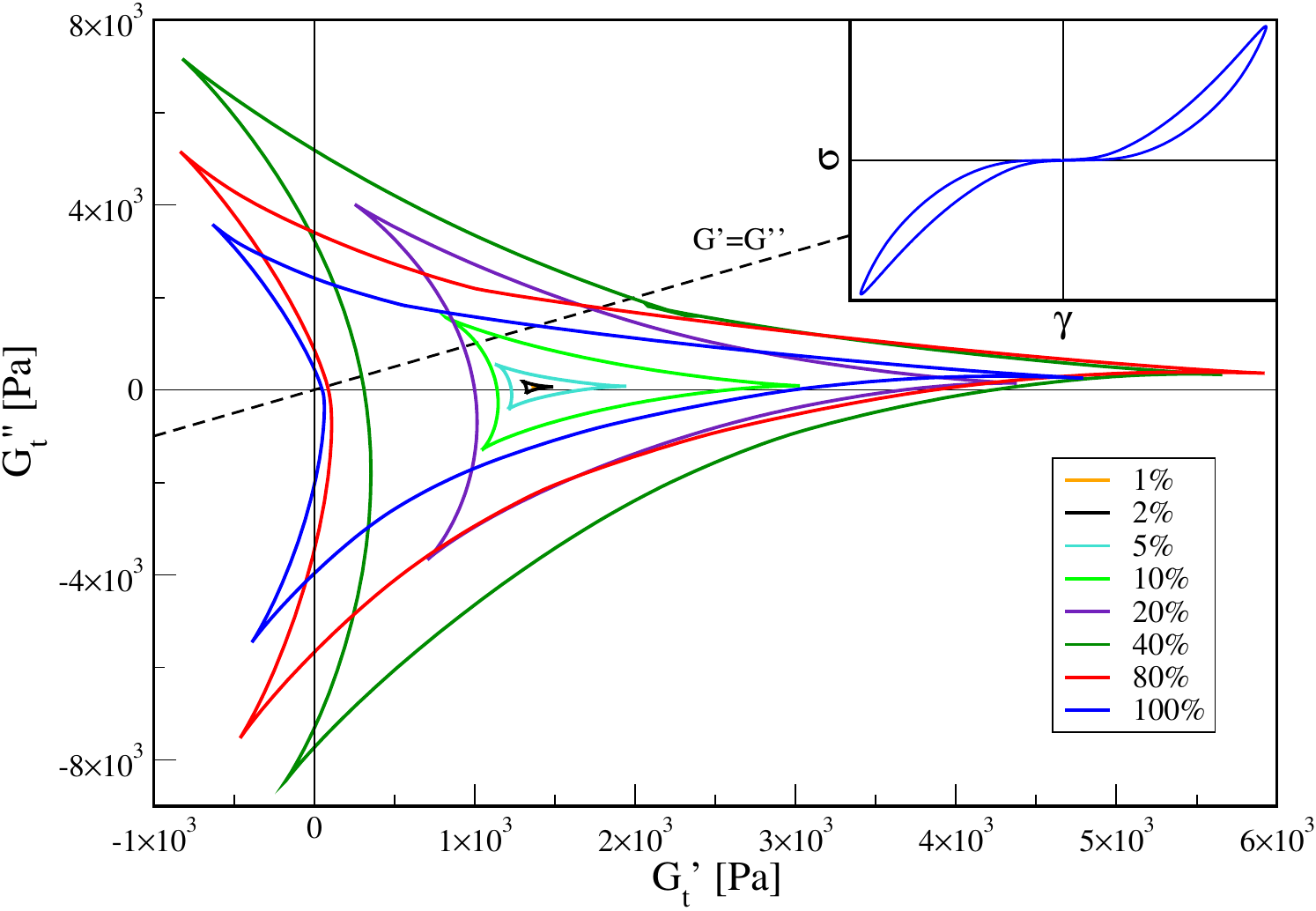}
    \caption{Cole-Cole plot of $G_t'$ and $G_t''$ for stress strain rate response curves at 0.2 \si{\hertz} for strain amplitudes ranging from 1\% to 100\% for the fibrin sub-network. In the top right hand corner, the shape of the stress strain response curve is plotted for a strain amplitude of 100\%.}
    \label{fig:SPP_fibrin}
\end{figure}

A typical example of a raw LAOS data for fibrin without transglutaminaise (F) ($\gamma_0 = 100\%$ ) is plotted in the inset of Figure \ref{fig:SPP_fibrin}, providing insights into the features of the SPP analysis for fibrin. Hysteresis and broadening of the stress-strain Lissajous curves indicate higher dissipation at large strains, while a plateau at low strains, centered about zero stress, implies that the gel has experienced plastic deformation often associated with fibril stretching  \cite{PIECHOCKA20102281}. The resulting SPP analysis provides several characteristics that further distinguish fibrin from gelatin; such as a larger overall deltoid area, broken symmetry for high strain amplitudes, shifting of the deltoid centers closer to the $G_t'=G_t''$ separatrix, and retraction of the apices at higher $\gamma_0$ (see Figure \ref{fig:SPP_fibrin}) -- (F+) gels show similar behavior (see supplemental materials). We will elucidate these features using the geometric analysis described above, providing a straight forward comparison to the gelatin gels.  

\begin{figure*}[htpb]
    \centering
    \includegraphics[width=1.00\linewidth]{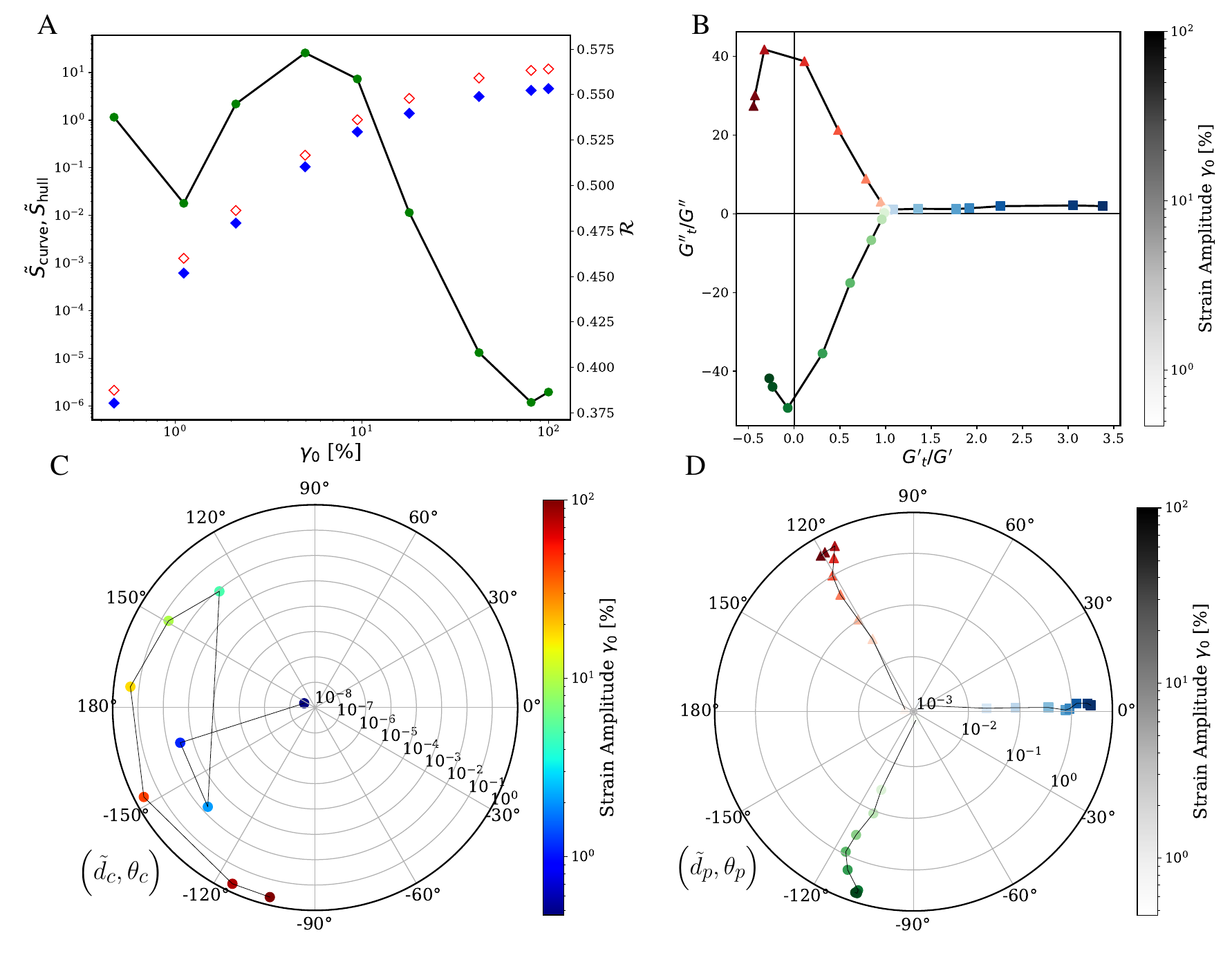}
    \caption{Evolution of the normalized measurements and values described in fig.\ref{fig:spp_legend} for the (F) subgel composition (fibrin only) with the strain amplitude.\textbf{(A)} Normalized areas of the parametric response, the convex hull (open red symbols) and parametric plot area (blue symbols) to hull area ratio (green symbols) as defined in fig.\ref{fig:spp_legend}. As the strain amplitude increases, we notice the gradual increase of both the the parametric plot and hull areas from values $\sim 10^{-6}$ at $0.5\%$ to $\sim 10^{1}$ at $100\%$ as well as a decrease of their ratio from $\sim 0.55$ to $\sim 0.4$, values that differ significantly from the values of the (G+) composition.\textbf{(B)} Normalized storage and loss moduli of the apices $\mathbf{\tilde{A}}_p$ as a function of the strain amplitude of the oscillatory shear imposed to the sample. The evolution of the position of the apices with strain amplitude can be observed and leads to an increasing distance between each apex and the increasing extension of the parametric plot. We also notice the change in trajectory of apices 2 and 3 as the materials at strain amplitude above $\gamma_0 = 40\%$, indicating significant changes in the overall rheological behavior of the (F) subgel. \textbf{(C)} Polar plot of the coordinates $\left[ \tilde{d}_c , \theta _c \right]$ of the vector between the rheological center ($\mathbf{C}_{\mathrm{R}}$) and the hull center ($\mathbf{C}_\mathrm{H}$) as defined in fig.\ref{fig:spp_legend}. We can observe the increasing distance between the rheological centers and hull centers as the strain amplitude increases. The norm increases with the strain amplitude applied from $\sim 10^{-8}$ at $\gamma_0 = 0.5\%$ to $\sim 10^{0}$ at $\gamma_0 = 100\%$, with the maximum value 2 orders of magnitude above the ones displayed by the (G+) composition. The values of the angles display a evolution varying from 130\degree at $\gamma_0 = 5\%$ to -90\degree at $\gamma_0 = 100\%$ indicating significant changes in the temporal dominant response of the (F) composition. As strain amplitude increases, the material transitions from a strain stiffening domininant behavior to a dissipative one at the highest strain amplitudes tested. \textbf{(D)} Plots of the coordinates of the vectors from the rheological center to each individual apex as described in fig.\ref{fig:spp_legend} for (F) subgel. We can observe the increasing distance between the rheological center and the apices as the strain amplitude increases from $\gamma_0=0.5\%$ to 40\% and a plateauing of this value above 40\%. The angles are distributed at angles 0\degree, 120\degree and -120\degree before $\gamma_0 = 40\%$ and exhibit $\sim$ 5$\degree$ to 10$\degree$ for apices 2 and 3 above this value. The evolutions of both the norm and angle show considerable differences from the ones from the (G+) subgel due to the change of rheological response occurring after $\gamma _0 > 40\%$}
    \label{fig:spp_analyzed_fibrin_unified}
\end{figure*}

In Figure \ref{fig:spp_analyzed_fibrin_unified}A $(\tilde{S}_{\rm curve}, \tilde{S}_{\rm hull}, {\cal R})$ are again plotted as a function of strain for the fibrin gels. Fibrin exhibits a marked difference in the magnitude and growth of $\tilde{S}_{\rm curve}$ as a function of strain compared to gelatin. The normalized areas exhibit a strong power-law increase for strains below $\gamma_0 = 20\%$, which then turns over to a plateau at large strain amplitudes. Moreover, the magnitude of the normalized areas reaches a value of nearly two orders of magnitude larger than that of gelatin.  The plateau corresponds to the decrease in the normalized area inscribed in the parametric plots at strain amplitudes larger than $\gamma _0 = 40\%$ observed in fig.\ref{fig:SPP_fibrin}. The origin of the plateau is explained by observing a decrease in the position of the rheological center along the $G'$ axis; indicating a reduction in the elastic response brought on by plastic deformations to the gel. We also observe that the area ratio for the fibrin gels decreases with increasing amplitude more quickly than that of the gelatin as a result of the increased concavity of the deltoid shape. Physically, this suggests that the nonlinearity of the response extremes are particularly pronounced compared to the overall response.

The normalized position of the apices in the $\left( G',G'' \right)$ space are shown in Figure \ref{fig:spp_analyzed_fibrin_unified}A. We observe in the positions the decrease of the values for $\left| G' \right|$ for apices 2 and 3 after $\gamma _0 = 40\%$ and the reverse motion of apex 1 after $\gamma _0 = 80\%$. This observation is consistent with the permanent structural modification occurring in the fibrin network at high strain amplitudes. Of particular note, the position of apex 2 (red data) crosses the $G'_t=G''_t$ separatrix at large amplitudes, indicating that there are portions of the period in the vicinity of that apex where viscous effects dominate, as well as the presence of yielding/unyielding transitions\cite{Donley2019,Donley2019_2}.

As observed for gelatin, the SPP-deltoid position for fibrin also encodes for deviations from linear viscoelasticity and the relative magnitude of the elastic to viscous responses, respectively. The polar coordinates $\left[ d_\mathrm{c}, \theta_\mathrm{c} \right]$ of the vector $\mathbf{\underline{C}}$ described in Section \ref{subsubsec:legend} are plotted for fibrin gels in Figure \ref{fig:spp_analyzed_fibrin_unified}C  for applied strains ranging from $\gamma _0 = 0.5\to 100\%$. We observe very different behavior compared to the gelatin results. First, we find that the magnitude of $d_c$ increases rapidly to values that are two orders of magnitude greater than that of gelatin as a function of applied strain, revealing a shift in time evolution of the SPP response resulting from a larger contribution from viscous dissipation. The value of $\theta_c$ remains close to $180\degree$ for low strain values but rapidly deviates and continuously varies from $130\degree \to -100\degree$. We again attribute this behavior to the transition from a time balanced response at low strain to a localization near apex 2 at higher strains, reinforcing the argument that plastic and dissipative behavior dominates at large strains. In Figure \ref{fig:spp_analyzed_fibrin_unified}D we plot the evolution of the polar coordinates $[d_p,\theta_c]$ of the vectors between $R_c$ and each apex $\mathbf{A}_p$ and again observe behavior distinguishable from gelatin. The apex angles $\theta_c$ remain largely centered on $[0\degree, \pm120\degree]$, but the lengths $d_p$ begin to converge to a constant value with minor changes in $\theta_c$ as the strain amplitude is increased. This symmetry breaking between apices 2 and 3 at large amplitudes also indicates that $G''_t$ is encoding some amount of viscous dissipation (as opposed to the instantaneous non-linear elasticity in the G+ case), and that the fibrin gel, unlike the gelatin network, is not fully elastic in this range.

\paragraph{Complete composite network: GF+}

\begin{figure}[htb]
    \centering
    \includegraphics[width=1.00\linewidth]{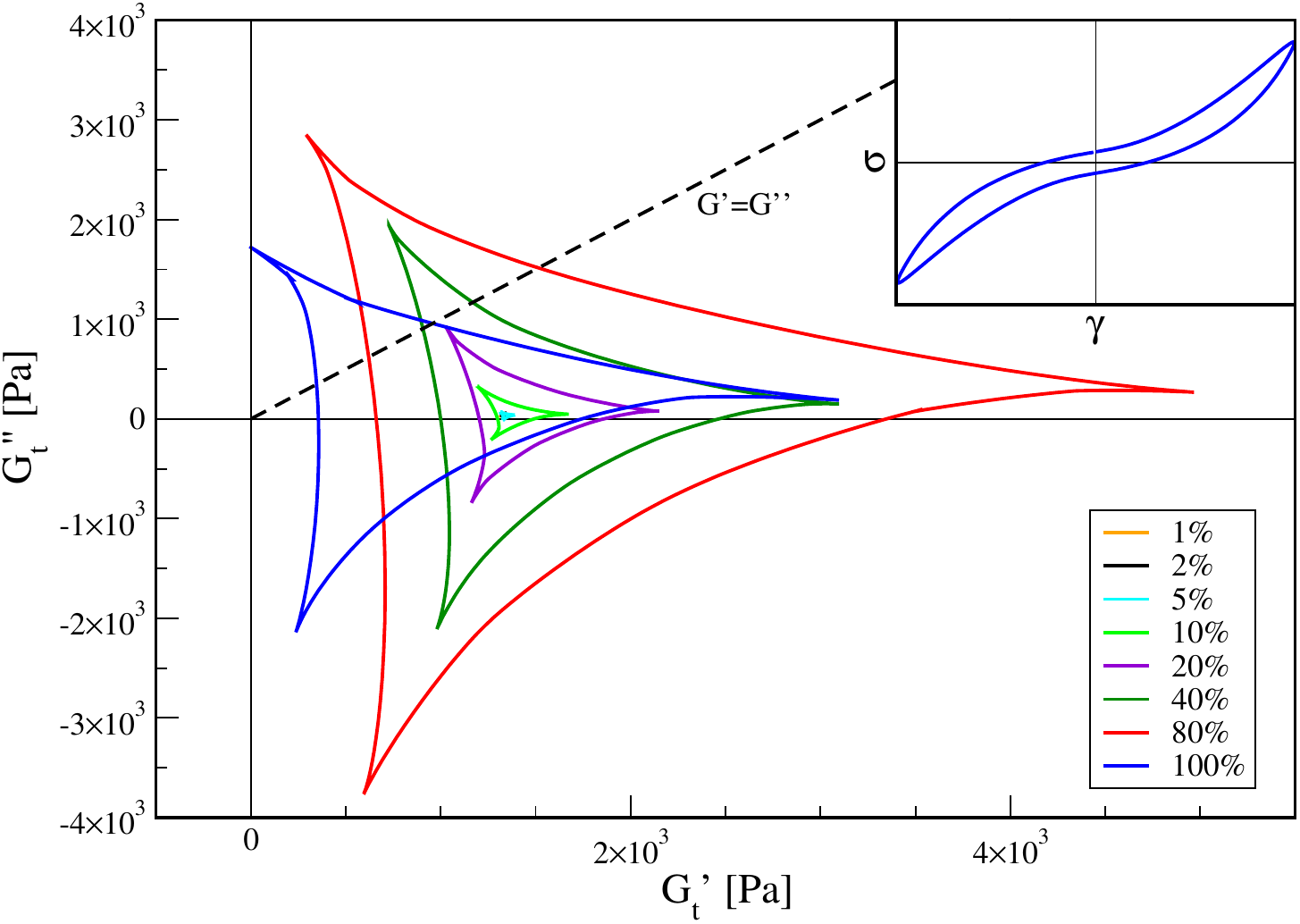}
    \caption{Cole-Cole plot of $G_t'$ and $G_t''$ for stress strain rate response curves at 0.2 \si{\hertz} for strain amplitudes ranging from 1\% to 100\% for the complete composite network. In the top right hand corner, the shape of the stress strain response curve is plotted for a strain amplitude of 100\%. The dashed line corresponds to $G_t' = G_t''$. We can observe the onset and increase in nonlinear behavior with the increasing area defined by the deltoids as well as their spatial extension in the $G_t', G_t''$ space. We can also note the decrease in spatial extension and area of the deltoids from $\gamma_0 = 80$\% to $\gamma_0 = 100$\%}
    \label{fig:SPP_full_gel}
\end{figure}

In this final section we describe the LAOS rheology and SPP analysis of the composite gel (GF+) composed of gelatin, fibrin and  transglutaminaise. As described in the Methods, the concentration of the (GF+) gel set to be the total weight percent of each independent gel described in the previous sections. We ensure phase separation is not occurring by monitoring the turbidity of the composite network during the co-polymerization process (data not shown). A typical LAOS rheology Lissajous curve for $\gamma_0 = 100\%$ is shown in Figure \ref{fig:SPP_full_gel} {\em inset}. It's noteworthy that the stress-strain response has some features of the fibrin gel, such as strong nonlinearity at high strains with a hysteretic broadening  of the curve, but is broadly lacking a dominant elastic response found in the crosslinked gelatin gels. The overall shape of the (GF+) SPP-deltoids is more reminiscent to the fibrin gels, owing to the lack of symmetry with the shifting 
of $\mathbf{A}_2$ toward the $G'_t$ axis and to lower values of $G'_t$ as compared to 
$\mathbf{A}_3$ for strain amplitudes $\gamma _0 \ge 10\%$. This shift is indicative of 
the onset of plastic deformations in the combined gel, 
which is more consistent with the microstructural changes observed for fibrin at high strains. While the similarities to the rheology of the fibrin gel are clear, 
a few notable differences 
are evident in the response of the combined gel. Specifically, the deltoids for the (GF+) remain largely below the $G'=G''$ separatrix, indicating that for $\gamma_0 \geq 40\%$ the response is dominated by $G'_t>G''_t$. This corresponds to a more dominant component of the overall response being controlled by elasticity rather than the yielding observed in fibrin. The persistence of elasticity may be attributed to the crosslinked gelatin network providing a elastic sub-network structure that remains once the fibrin microstructure irreversibly deforms at higher strains. 

\begin{figure*}[htpb]
    \centering
    \includegraphics[width=1.00\linewidth]{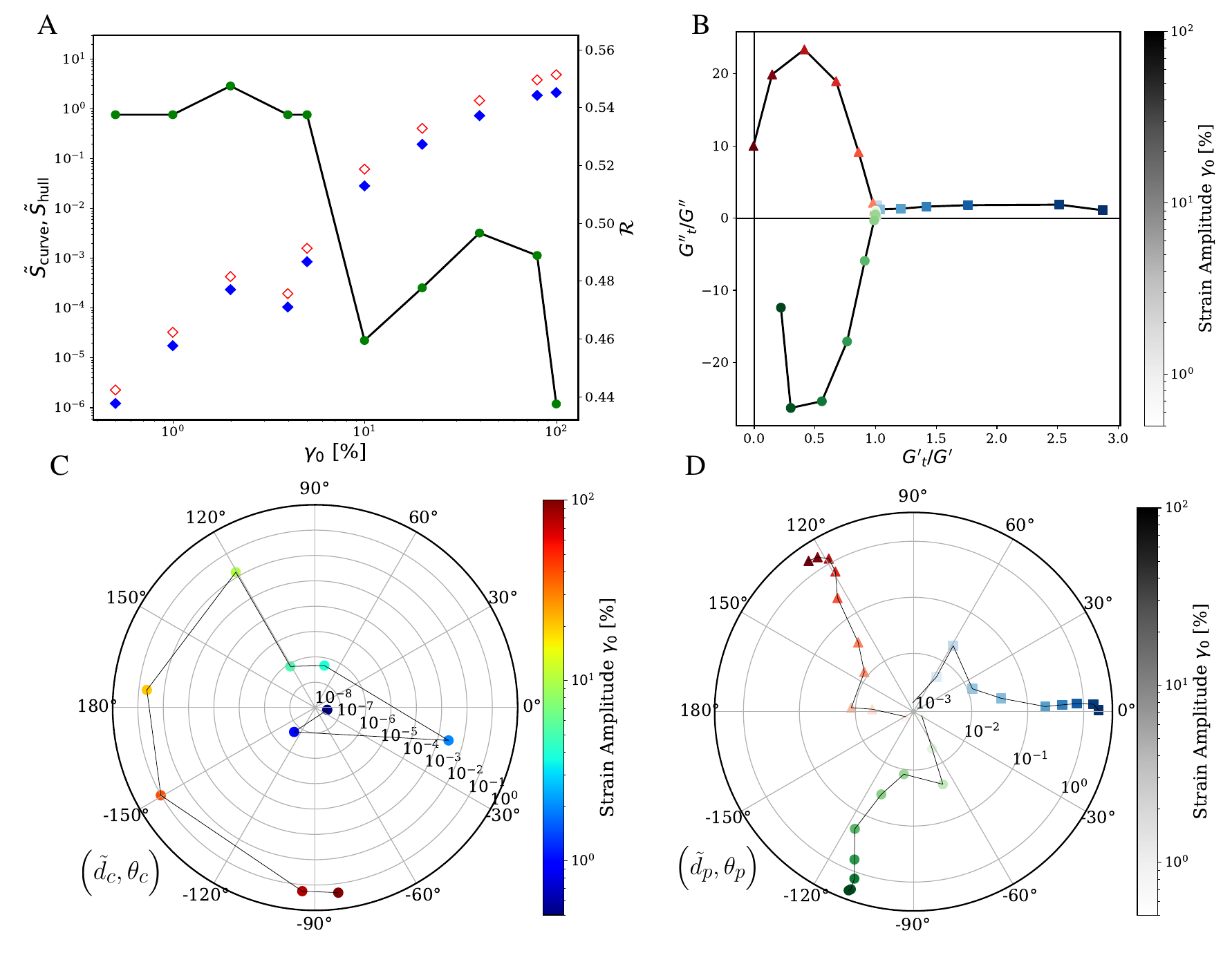}
    \caption{Evolution of the normalized measurements and values described in fig.\ref{fig:spp_legend} for the (GF+) composition (full composite gel) with the strain amplitude. \textbf{(A)} Normalized areas of the parametric response, the convex hull (open red symbols) and parametric plot area (blue symbols) to hull area ratio (green ratios) as defined in fig.\ref{fig:spp_legend} for the complete composite gel ((GF+) composition). As the strain amplitude increases, we notice the gradual increase of both the the parametric plot and hull areas from values $\sim 10^{-6}$ at $0.5\%$ to $\sim 10^{0}$ at $100\%$ as well as a decrease of their ratio from $\sim 0.55$ to $\sim 0.45$, values comparable to the (F+) composition. \textbf{(B)}  Normalized storage and loss moduli of the apices $\mathbf{\tilde{A}}_p$ as a function of the strain amplitude of the oscillatory shear imposed to the sample. The evolution of the position of the apices with strain amplitude can be observed and leads to an increasing distance between each apex and the increasing extension of the parametric plot. We also notice the change in trajectory of apices 2 and 3 as the materials experiences changes in mechanical properties after $\gamma_0 = 40\%$. \textbf{(C)} Polar plot of the coordinates $\left[ \tilde{d}_c , \theta _c \right]$ of the vector between $\mathbf{C}_{\mathrm{R}}$ and $\mathbf{C}_{\mathrm{H}}$ as defined in fig.\ref{fig:spp_legend}. We can observe the increasing distance between the rheological centers and hull centers as the strain amplitude increases. The norm increases with the strain amplitude applied from $\sim 10^{-8}$ at $\gamma_0 = 0.5\%$ to $\sim 10^{-1}$ at $\gamma_0 = 100\%$. The values of the angles display a evolution varying from 90\degree at $\gamma_0 = 5\%$ to -80\degree at $\gamma_0 = 100\%$. \textbf{(D)} Plots of the coordinates of the vectors from the rheological center to each individual apex as described in fig.\ref{fig:spp_legend} for the complete composite gel. We can observe the increasing distance between the rheological center and the apices as the strain amplitude increases from $\gamma_0=0.5\%$ to 40\% and a plateauing of this value above 40\%. The angles are distributed at angles 0\degree, 120\degree and -120\degree before $\gamma_0 = 40\%$ and exhibit $\sim$ 5\degree to 10\degree for all apices above this value. The ``kink'' observed at $\gamma_0 = 2\%$ is an artefact of the SPP and our analysis framework.}
    \label{fig:spp_analyzed_fullgel_unified}
\end{figure*}

As with the constituents, we analyze the SPP-deltoids to extract principle geometric quantities of the (GF+) gels as a function of applied strain.  Perhaps unexpectedly, the $\tilde{S}_{\rm curve}, \tilde{S}_{\rm hull}$ values and their ratio $\cal{R}$ exhibit a strain dependence with features that combine behavior found in fibrin and gelatin gels while simultaneously preserving feature that are distinctive to the composite network (see Fig. \ref{fig:spp_analyzed_fullgel_unified}A). The normalized areas increase rapidly with a large positive power-law for strains below $\gamma_0<10\%$, followed by a rapid decrease at $\gamma_0=20\%$, followed by a recovery and plateau with a magnitude between that of fibrin and gelatin. The area ratio also exhibits a plateau at ${\cal R} \sim 0.54$, for $\gamma_0\le 20\%$ and then decays sharply and fluctuates as the strain is increased, finally decaying to a value below that of fibrin. These low ${\cal R}$ values are indicative of a change in concavity of the SPP-deltoids as the scale of the nonlinearity increases. 

The normalized magnitude of the apex values for the (GF+) gels shown in Figure \ref{fig:spp_analyzed_fullgel_unified}B. As with both fibrin and gelatin, the relative magnitude of $\mathbf{A}_1$ continues to increase over all strains, whereas $\mathbf{A}_{2,3}$ begin to retract beyond $\gamma_0>20\%$, similar to fibrin. Moreover, the growth of $\mathbf{A}_1$ remains flat along the the $G'_t/G'$ axis and maintains a similar magnitude to the fibrin case. However, for $\mathbf{A}_{2,3}$, the growth along the $G''_t/G''$ axis is a factor of 2 lower than that of fibrin prior to the reduction beyond the peak at $\gamma_0=40\%$. It's also noteworthy that the value of all $\mathbf{A}_p$ values never drop below $G'_t/G' =0$ for all strain. The evolution of $[\tilde{d}_c,\theta_c]$ for the (GF+) gels follow nearly identical angular values and magnitudes as fibrin for all $\gamma_0$, with a slight deviation at $\gamma_0=5.0\%$. We also quantify the normalized vector and orientation $[\tilde{d}_p,\theta_p]$ connecting the $R_c$ to each $\mathbf{A}_p$ for the (GF+) gels. Here we also observe similarities, both quantitative and phenomenological to that of the fibrin gels, noting that distinctions remain at $\gamma_0=5.0\%$ (which may be an artifact of the analysis) and for the angular dependence for $\mathbf{A}_3$ (see Fig. \ref{fig:spp_analyzed_fullgel_unified}D). 

\section{Conclusions}
\label{conclusions}

Bio-compatible composite gels are becoming critically important materials for a variety of applications ranging from tumor scaffolds, three-dimensional cell culture and bio-printing. Each use requires a strict set of physical and homeostatic criteria that are vital to ensure cellular viability. Many of the processing conditions found in extrusion 3D-printing require the imposition of large strains at high rates, resulting in conditions that can ultimately lead to reduced cell survival in the final material. Moreover, polymerization kinetics of the sub-networks must also be tuned to optimize the flowabilty and solidification criteria for the particular application. In many cases, the strains experienced by the composite gels can exceed the typical yield values found in many biopolymer networks, while for gels composed of fibrin, the effect of large strains is often an enhancement of nonlinearity in the mechanical behavior such as strain stiffening and plasticity of the final network architecture. Conventional rheological methods that probe the linear response are, by definition, unable to explore the nonlinear rheology found when materials reach their yield criteria. 

Using Large Amplitude Oscillatory (LAOS) rheology we have measured the mechanical properties of prototypical single and composite biopolymer gels composed of gelatin and fibrin that are crosslinked with transglutaminaise. By building on 
the Sequence of Physical Processes framework (SPP) \cite{Rogers2017}, we have developed 
a set of statistical and geometrical tools that provide a robust platform for extracting relevant and rheologically distinct parameters that allow for the direct comparison of the nonlinear rheology of the biocomposite and its constituents. These new tools provide the the means for a facile comparison of LAOS rheological data across complex composite materials -- a highly relevant, but underdeveloped tool-set.

The application of the Sequence of Physical Processes (SPP) framework to LAOS data allows for 
a comprehensive, time-resolved understanding of the 
oscillatory stress-strain responses of the composite gels and their components, through the time-dependent moduli $G'_t$ and $G''_t$. 
As these are instantaneous values 
derived at 
every point of the oscillatory responses, the amount of data and information contained in the temporal moduli values remains identical to the raw Lissajous curves. In order to facilitate comparisons across systems, 
we have extended the SPP-framework into a systematic analysis of the shapes of the time-dependent Cole-Cole plots, which reduces the SPP results into a set of parameters that are physically and rheologically meaningful and can provide quantitative insights often gleaned only qualitatively from conventional LAOS Lissajous curves. 
The form of the SPP-deltoids is quantified by the relative area of the enclosing hull to determine the strain dependence of the deltoid concavity, which indicates the transition away from linear response. Taken together with the orientation of the apices, we can determine if the shift in hull geometry is indicative of stiffening or dissipation within the network by considering the preservation of the overall deltoid symmetry.

Our geometric analysis shows that crosslinked gelatin gels exhibit highly elastic behavior that transitions to stiffening at high strain, whereas over the same range of strain, fibrin exhibits an increase in dissipation followed by strain induced alterations to the microstructure. The same analysis also provides clearly identifiable regimes where the combined networks influence the rheology of the composite gel. Interestingly, we observe that the (GF+) gels are not simply a linear combination of gelatin and fibrin, resulting in an emergent and {\em non-additive} rheology that preserves specific strain dependent behavior found within each sub-gel.

 The analysis proposed here also shows that the dissipative behavior as well as the permanent microstructural deformations at higher strain amplitude present is present in the (GF+) gels which is similar to the fibrin subgel. Nonetheless, the changes of the time-resolved mechanical responses due to the permanent structural modifications are less pronounced in the (GF+) composite network, suggesting that the gelatin network plays an important role through its interactions with the fibrin network that are perhaps mediated by the sharing of transglutaminase crosslinkers. The complete composite network exhibits hysteresis over the whole cycle, with a non zero stress and ${\partial \sigma}/{\partial \gamma}$ slope at strain of 0\%, where the fibrin network shows a zero stress and a zero ${\partial \sigma}/{\partial \gamma}$ slope under the same conditions, indicating that gelatin is imparting residual elasticity to the network structure.

Additionally, as characterization and quick fingerprinting of nonlinear responses in composite materials is a difficult undertaking, the statistical and geometrical parameters derived from our analysis of the time-resolved rheology, which distill key physically relevant characteristics of the nonlinear response of soft materials under high strain, could be used in conjunction with machine learning (ML) models to uncover hidden statistical information and help categorize main nonlinear oscillatory behaviors. 



\section{Acknowledgments}

\section{Author Contributions}
Conceptualization: Methodology: Investigation: Visualization: Writing:  Editing: Funding Acquisition: Supervision: . 

\section{Author Competing Interests}
None.


%
%

%


\bibliography{localbibliography.bib}

\begin{thebibliography}{45}%
\makeatletter
\providecommand \@ifxundefined [1]{%
 \@ifx{#1\undefined}
}%
\providecommand \@ifnum [1]{%
 \ifnum #1\expandafter \@firstoftwo
 \else \expandafter \@secondoftwo
 \fi
}%
\providecommand \@ifx [1]{%
 \ifx #1\expandafter \@firstoftwo
 \else \expandafter \@secondoftwo
 \fi
}%
\providecommand \natexlab [1]{#1}%
\providecommand \enquote  [1]{``#1''}%
\providecommand \bibnamefont  [1]{#1}%
\providecommand \bibfnamefont [1]{#1}%
\providecommand \citenamefont [1]{#1}%
\providecommand \href@noop [0]{\@secondoftwo}%
\providecommand \href [0]{\begingroup \@sanitize@url \@href}%
\providecommand \@href[1]{\@@startlink{#1}\@@href}%
\providecommand \@@href[1]{\endgroup#1\@@endlink}%
\providecommand \@sanitize@url [0]{\catcode `\\12\catcode `\$12\catcode
  `\&12\catcode `\#12\catcode `\^12\catcode `\_12\catcode `\%12\relax}%
\providecommand \@@startlink[1]{}%
\providecommand \@@endlink[0]{}%
\providecommand \url  [0]{\begingroup\@sanitize@url \@url }%
\providecommand \@url [1]{\endgroup\@href {#1}{\urlprefix }}%
\providecommand \urlprefix  [0]{URL }%
\providecommand \Eprint [0]{\href }%
\providecommand \doibase [0]{https://doi.org/}%
\providecommand \selectlanguage [0]{\@gobble}%
\providecommand \bibinfo  [0]{\@secondoftwo}%
\providecommand \bibfield  [0]{\@secondoftwo}%
\providecommand \translation [1]{[#1]}%
\providecommand \BibitemOpen [0]{}%
\providecommand \bibitemStop [0]{}%
\providecommand \bibitemNoStop [0]{.\EOS\space}%
\providecommand \EOS [0]{\spacefactor3000\relax}%
\providecommand \BibitemShut  [1]{\csname bibitem#1\endcsname}%
\let\auto@bib@innerbib\@empty
\bibitem [{\citenamefont {Rogers}(2017)}]{Rogers2017}%
  \BibitemOpen
  \bibfield  {author} {\bibinfo {author} {\bibfnamefont {S.~A.}\ \bibnamefont
  {Rogers}},\ }\bibfield  {title} {\enquote {\bibinfo {title} {In search of
  physical meaning: defining transient parameters for nonlinear
  viscoelasticity},}\ }\href {https://doi.org/10.1007/s00397-017-1008-1}
  {\bibfield  {journal} {\bibinfo  {journal} {Rheologica Acta}\ }\textbf
  {\bibinfo {volume} {56}},\ \bibinfo {pages} {501--525} (\bibinfo {year}
  {2017})}\BibitemShut {NoStop}%
\bibitem [{\citenamefont {Gungor-Ozkerim}\ \emph {et~al.}(2018)\citenamefont
  {Gungor-Ozkerim}, \citenamefont {Inci}, \citenamefont {Zhang}, \citenamefont
  {Khademhosseini},\ and\ \citenamefont
  {Dokmeci}}]{gungor-ozkerim_bioinks_2018}%
  \BibitemOpen
  \bibfield  {author} {\bibinfo {author} {\bibfnamefont {P.~S.}\ \bibnamefont
  {Gungor-Ozkerim}}, \bibinfo {author} {\bibfnamefont {I.}~\bibnamefont
  {Inci}}, \bibinfo {author} {\bibfnamefont {Y.~S.}\ \bibnamefont {Zhang}},
  \bibinfo {author} {\bibfnamefont {A.}~\bibnamefont {Khademhosseini}},\ and\
  \bibinfo {author} {\bibfnamefont {M.~R.}\ \bibnamefont {Dokmeci}},\
  }\bibfield  {title} {\enquote {\bibinfo {title} {Bioinks for {3D}
  bioprinting: an overview.}}\ }\href@noop {} {\bibfield  {journal} {\bibinfo
  {journal} {Biomaterials science}\ }\textbf {\bibinfo {volume} {6}},\ \bibinfo
  {pages} {915--946} (\bibinfo {year} {2018})}\BibitemShut {NoStop}%
\bibitem [{\citenamefont {Rosales}\ and\ \citenamefont
  {Anseth}(2016)}]{rosales_design_2016}%
  \BibitemOpen
  \bibfield  {author} {\bibinfo {author} {\bibfnamefont {A.~M.}\ \bibnamefont
  {Rosales}}\ and\ \bibinfo {author} {\bibfnamefont {K.~S.}\ \bibnamefont
  {Anseth}},\ }\bibfield  {title} {\enquote {\bibinfo {title} {The design of
  reversible hydrogels to capture extracellular matrix dynamics},}\ }\href
  {https://doi.org/10.1038/natrevmats.2015.12} {\bibfield  {journal} {\bibinfo
  {journal} {Nature Reviews Materials}\ }\textbf {\bibinfo {volume} {1}},\
  \bibinfo {pages} {15012} (\bibinfo {year} {2016})}\BibitemShut {NoStop}%
\bibitem [{\citenamefont {Buchmann}, \citenamefont {Fernández},\ and\
  \citenamefont {Bausch}(2021)}]{Buchmann2021}%
  \BibitemOpen
  \bibfield  {author} {\bibinfo {author} {\bibfnamefont {B.}~\bibnamefont
  {Buchmann}}, \bibinfo {author} {\bibfnamefont {P.}~\bibnamefont
  {Fernández}},\ and\ \bibinfo {author} {\bibfnamefont {A.~R.}\ \bibnamefont
  {Bausch}},\ }\bibfield  {title} {\enquote {\bibinfo {title} {The role of
  nonlinear mechanical properties of biomimetic hydrogels for organoid
  growth},}\ }\href {https://doi.org/10.1063/5.0044653} {\bibfield  {journal}
  {\bibinfo  {journal} {Biophysics Reviews}\ }\textbf {\bibinfo {volume} {2}}
  (\bibinfo {year} {2021}),\ 10.1063/5.0044653}\BibitemShut {NoStop}%
\bibitem [{\citenamefont {Kim}\ \emph {et~al.}(2022)\citenamefont {Kim},
  \citenamefont {Lee}, \citenamefont {Kotula}, \citenamefont {Takagi},
  \citenamefont {Chow},\ and\ \citenamefont {Alimperti}}]{Kim2022}%
  \BibitemOpen
  \bibfield  {author} {\bibinfo {author} {\bibfnamefont {Y.}~\bibnamefont
  {Kim}}, \bibinfo {author} {\bibfnamefont {E.-J.}\ \bibnamefont {Lee}},
  \bibinfo {author} {\bibfnamefont {A.~P.}\ \bibnamefont {Kotula}}, \bibinfo
  {author} {\bibfnamefont {S.}~\bibnamefont {Takagi}}, \bibinfo {author}
  {\bibfnamefont {L.}~\bibnamefont {Chow}},\ and\ \bibinfo {author}
  {\bibfnamefont {S.}~\bibnamefont {Alimperti}},\ }\bibfield  {title} {\enquote
  {\bibinfo {title} {Engineering 3d printed scaffolds with tunable
  hydroxyapatite},}\ }\href {https://doi.org/10.3390/jfb13020034} {\bibfield
  {journal} {\bibinfo  {journal} {Journal of Functional Biomaterials}\ }\textbf
  {\bibinfo {volume} {13}},\ \bibinfo {pages} {34} (\bibinfo {year}
  {2022})}\BibitemShut {NoStop}%
\bibitem [{\citenamefont {Wang}\ \emph {et~al.}(2017)\citenamefont {Wang},
  \citenamefont {Ao}, \citenamefont {Tian}, \citenamefont {Fan}, \citenamefont
  {Tong}, \citenamefont {Hou},\ and\ \citenamefont {Bai}}]{Wang2017}%
  \BibitemOpen
  \bibfield  {author} {\bibinfo {author} {\bibfnamefont {X.}~\bibnamefont
  {Wang}}, \bibinfo {author} {\bibfnamefont {Q.}~\bibnamefont {Ao}}, \bibinfo
  {author} {\bibfnamefont {X.}~\bibnamefont {Tian}}, \bibinfo {author}
  {\bibfnamefont {J.}~\bibnamefont {Fan}}, \bibinfo {author} {\bibfnamefont
  {H.}~\bibnamefont {Tong}}, \bibinfo {author} {\bibfnamefont {W.}~\bibnamefont
  {Hou}},\ and\ \bibinfo {author} {\bibfnamefont {S.}~\bibnamefont {Bai}},\
  }\bibfield  {title} {\enquote {\bibinfo {title} {Gelatin-based hydrogels for
  organ 3d bioprinting},}\ }\href {https://doi.org/10.3390/polym9090401}
  {\bibfield  {journal} {\bibinfo  {journal} {Polymers}\ }\textbf {\bibinfo
  {volume} {9}},\ \bibinfo {pages} {401} (\bibinfo {year} {2017})}\BibitemShut
  {NoStop}%
\bibitem [{\citenamefont {Mierke}\ \emph {et~al.}(2011)\citenamefont {Mierke},
  \citenamefont {Frey}, \citenamefont {Fellner}, \citenamefont {Herrmann},\
  and\ \citenamefont {Fabry}}]{Mierke2011}%
  \BibitemOpen
  \bibfield  {author} {\bibinfo {author} {\bibfnamefont {C.~T.}\ \bibnamefont
  {Mierke}}, \bibinfo {author} {\bibfnamefont {B.}~\bibnamefont {Frey}},
  \bibinfo {author} {\bibfnamefont {M.}~\bibnamefont {Fellner}}, \bibinfo
  {author} {\bibfnamefont {M.}~\bibnamefont {Herrmann}},\ and\ \bibinfo
  {author} {\bibfnamefont {B.}~\bibnamefont {Fabry}},\ }\bibfield  {title}
  {\enquote {\bibinfo {title} {Integrin $\alpha$5$\beta$1 facilitates cancer
  cell invasion through enhanced contractile forces},}\ }\href
  {https://doi.org/10.1242/jcs.071985} {\bibfield  {journal} {\bibinfo
  {journal} {Journal of Cell Science}\ }\textbf {\bibinfo {volume} {124}},\
  \bibinfo {pages} {369--383} (\bibinfo {year} {2011})}\BibitemShut {NoStop}%
\bibitem [{\citenamefont {Kai}, \citenamefont {Laklai},\ and\ \citenamefont
  {Weaver}(2016)}]{Kai2016}%
  \BibitemOpen
  \bibfield  {author} {\bibinfo {author} {\bibfnamefont {F.}~\bibnamefont
  {Kai}}, \bibinfo {author} {\bibfnamefont {H.}~\bibnamefont {Laklai}},\ and\
  \bibinfo {author} {\bibfnamefont {V.~M.}\ \bibnamefont {Weaver}},\ }\bibfield
   {title} {\enquote {\bibinfo {title} {Force matters: Biomechanical regulation
  of cell invasion and migration in disease},}\ }\href
  {https://doi.org/10.1016/j.tcb.2016.03.007} {\bibfield  {journal} {\bibinfo
  {journal} {Trends in Cell Biology}\ }\textbf {\bibinfo {volume} {26}},\
  \bibinfo {pages} {486--497} (\bibinfo {year} {2016})}\BibitemShut {NoStop}%
\bibitem [{\citenamefont {Parsons}, \citenamefont {Horwitz},\ and\
  \citenamefont {Schwartz}(2010)}]{Parsons2010}%
  \BibitemOpen
  \bibfield  {author} {\bibinfo {author} {\bibfnamefont {J.~T.}\ \bibnamefont
  {Parsons}}, \bibinfo {author} {\bibfnamefont {A.~R.}\ \bibnamefont
  {Horwitz}},\ and\ \bibinfo {author} {\bibfnamefont {M.~A.}\ \bibnamefont
  {Schwartz}},\ }\bibfield  {title} {\enquote {\bibinfo {title} {Cell adhesion:
  integrating cytoskeletal dynamics and cellular tension},}\ }\href
  {https://doi.org/10.1038/nrm2957} {\bibfield  {journal} {\bibinfo  {journal}
  {Nature Reviews Molecular Cell Biology}\ }\textbf {\bibinfo {volume} {11}},\
  \bibinfo {pages} {633--643} (\bibinfo {year} {2010})}\BibitemShut {NoStop}%
\bibitem [{\citenamefont {Rijns}\ \emph {et~al.}(2024)\citenamefont {Rijns},
  \citenamefont {Rutten}, \citenamefont {Bellan}, \citenamefont {Yuan},
  \citenamefont {Mugnai}, \citenamefont {Rocha}, \citenamefont {Gado},
  \citenamefont {Kouwer},\ and\ \citenamefont {Dankers}}]{Rijns2024}%
  \BibitemOpen
  \bibfield  {author} {\bibinfo {author} {\bibfnamefont {L.}~\bibnamefont
  {Rijns}}, \bibinfo {author} {\bibfnamefont {M.~G. T.~A.}\ \bibnamefont
  {Rutten}}, \bibinfo {author} {\bibfnamefont {R.}~\bibnamefont {Bellan}},
  \bibinfo {author} {\bibfnamefont {H.}~\bibnamefont {Yuan}}, \bibinfo {author}
  {\bibfnamefont {M.~L.}\ \bibnamefont {Mugnai}}, \bibinfo {author}
  {\bibfnamefont {S.}~\bibnamefont {Rocha}}, \bibinfo {author} {\bibfnamefont
  {E.~D.}\ \bibnamefont {Gado}}, \bibinfo {author} {\bibfnamefont {P.~H.~J.}\
  \bibnamefont {Kouwer}},\ and\ \bibinfo {author} {\bibfnamefont {P.~Y.~W.}\
  \bibnamefont {Dankers}},\ }\bibfield  {title} {\enquote {\bibinfo {title}
  {Synthetic, multi-dynamic hydrogels by uniting stress-stiffening and
  supramolecular polymers},}\ }\href@noop {} {\bibfield  {journal} {\bibinfo
  {journal} {Science Advances}\ }\textbf {\bibinfo {volume} {10}},\ \bibinfo
  {pages} {eadr3209} (\bibinfo {year} {2024})}\BibitemShut {NoStop}%
\bibitem [{\citenamefont {Webber}\ \emph {et~al.}(2007)\citenamefont {Webber},
  \citenamefont {Creton}, \citenamefont {Brown},\ and\ \citenamefont
  {Gong}}]{Webber2007}%
  \BibitemOpen
  \bibfield  {author} {\bibinfo {author} {\bibfnamefont {R.~E.}\ \bibnamefont
  {Webber}}, \bibinfo {author} {\bibfnamefont {C.}~\bibnamefont {Creton}},
  \bibinfo {author} {\bibfnamefont {H.~R.}\ \bibnamefont {Brown}},\ and\
  \bibinfo {author} {\bibfnamefont {J.~P.}\ \bibnamefont {Gong}},\ }\bibfield
  {title} {\enquote {\bibinfo {title} {Large strain hysteresis and mullins
  effect of tough double-network hydrogels},}\ }\href
  {https://doi.org/10.1021/ma062924y} {\bibfield  {journal} {\bibinfo
  {journal} {Macromolecules}\ }\textbf {\bibinfo {volume} {40}},\ \bibinfo
  {pages} {2919--2927} (\bibinfo {year} {2007})}\BibitemShut {NoStop}%
\bibitem [{\citenamefont {Vereroudakis}\ \emph {et~al.}(2020)\citenamefont
  {Vereroudakis}, \citenamefont {Bantawa}, \citenamefont {Lafleur},
  \citenamefont {Parisi}, \citenamefont {Matsumoto}, \citenamefont {Peeters},
  \citenamefont {Del~Gado}, \citenamefont {Meijer},\ and\ \citenamefont
  {Vlassopoulos}}]{vereroudakis2020competitive}%
  \BibitemOpen
  \bibfield  {author} {\bibinfo {author} {\bibfnamefont {E.}~\bibnamefont
  {Vereroudakis}}, \bibinfo {author} {\bibfnamefont {M.}~\bibnamefont
  {Bantawa}}, \bibinfo {author} {\bibfnamefont {R.~P.~M.}\ \bibnamefont
  {Lafleur}}, \bibinfo {author} {\bibfnamefont {D.}~\bibnamefont {Parisi}},
  \bibinfo {author} {\bibfnamefont {N.~M.}\ \bibnamefont {Matsumoto}}, \bibinfo
  {author} {\bibfnamefont {J.~W.}\ \bibnamefont {Peeters}}, \bibinfo {author}
  {\bibfnamefont {E.}~\bibnamefont {Del~Gado}}, \bibinfo {author}
  {\bibfnamefont {E.~W.}\ \bibnamefont {Meijer}},\ and\ \bibinfo {author}
  {\bibfnamefont {D.}~\bibnamefont {Vlassopoulos}},\ }\bibfield  {title}
  {\enquote {\bibinfo {title} {Competitive supramolecular associations mediate
  the viscoelasticity of binary hydrogels},}\ }\href
  {https://doi.org/10.1021/acscentsci.0c00279} {\bibfield  {journal} {\bibinfo
  {journal} {ACS Central Science}\ }\textbf {\bibinfo {volume} {6}},\ \bibinfo
  {pages} {1401--1411} (\bibinfo {year} {2020})},\ \bibinfo {note} {pMID:
  32875081},\ \Eprint
  {https://arxiv.org/abs/https://doi.org/10.1021/acscentsci.0c00279}
  {https://doi.org/10.1021/acscentsci.0c00279} \BibitemShut {NoStop}%
\bibitem [{\citenamefont {Mugnai}, \citenamefont {Batoum},\ and\ \citenamefont
  {Gado}(2025)}]{mugnai2025interspecies}%
  \BibitemOpen
  \bibfield  {author} {\bibinfo {author} {\bibfnamefont {M.~L.}\ \bibnamefont
  {Mugnai}}, \bibinfo {author} {\bibfnamefont {R.~T.}\ \bibnamefont {Batoum}},\
  and\ \bibinfo {author} {\bibfnamefont {E.~D.}\ \bibnamefont {Gado}},\
  }\bibfield  {title} {\enquote {\bibinfo {title} {Interspecies interactions in
  dual, fibrous gels enable control of gel structure and rheology},}\ }\href
  {https://doi.org/10.1073/pnas.2423293122} {\bibfield  {journal} {\bibinfo
  {journal} {Proceedings of the National Academy of Sciences}\ }\textbf
  {\bibinfo {volume} {122}},\ \bibinfo {pages} {e2423293122} (\bibinfo {year}
  {2025})},\ \Eprint
  {https://arxiv.org/abs/https://www.pnas.org/doi/pdf/10.1073/pnas.2423293122}
  {https://www.pnas.org/doi/pdf/10.1073/pnas.2423293122} \BibitemShut {NoStop}%
\bibitem [{\citenamefont {Roshanbinfar}\ \emph {et~al.}(2025)\citenamefont
  {Roshanbinfar}, \citenamefont {Evans}, \citenamefont {Samanta}, \citenamefont
  {Kolesnik-Gray}, \citenamefont {Fiedler}, \citenamefont {Krstic},
  \citenamefont {Engel},\ and\ \citenamefont
  {Oommen}}]{ROSHANBINFAR2025123174}%
  \BibitemOpen
  \bibfield  {author} {\bibinfo {author} {\bibfnamefont {K.}~\bibnamefont
  {Roshanbinfar}}, \bibinfo {author} {\bibfnamefont {A.~D.}\ \bibnamefont
  {Evans}}, \bibinfo {author} {\bibfnamefont {S.}~\bibnamefont {Samanta}},
  \bibinfo {author} {\bibfnamefont {M.}~\bibnamefont {Kolesnik-Gray}}, \bibinfo
  {author} {\bibfnamefont {M.}~\bibnamefont {Fiedler}}, \bibinfo {author}
  {\bibfnamefont {V.}~\bibnamefont {Krstic}}, \bibinfo {author} {\bibfnamefont
  {F.~B.}\ \bibnamefont {Engel}},\ and\ \bibinfo {author} {\bibfnamefont
  {O.~P.}\ \bibnamefont {Oommen}},\ }\bibfield  {title} {\enquote {\bibinfo
  {title} {Enhancing biofabrication: Shrink-resistant collagen-hyaluronan
  composite hydrogel for tissue engineering and 3d bioprinting applications},}\
  }\href {https://doi.org/https://doi.org/10.1016/j.biomaterials.2025.123174}
  {\bibfield  {journal} {\bibinfo  {journal} {Biomaterials}\ }\textbf {\bibinfo
  {volume} {318}},\ \bibinfo {pages} {123174} (\bibinfo {year}
  {2025})}\BibitemShut {NoStop}%
\bibitem [{\citenamefont {Rijns}, \citenamefont {Baker},\ and\ \citenamefont
  {Dankers}(2024)}]{Rijns2024JACS}%
  \BibitemOpen
  \bibfield  {author} {\bibinfo {author} {\bibfnamefont {L.}~\bibnamefont
  {Rijns}}, \bibinfo {author} {\bibfnamefont {M.~B.}\ \bibnamefont {Baker}},\
  and\ \bibinfo {author} {\bibfnamefont {P.~Y.~W.}\ \bibnamefont {Dankers}},\
  }\bibfield  {title} {\enquote {\bibinfo {title} {Using chemistry to recreate
  the complexity of the extracellular matrix: Guidelines for supramolecular
  hydrogel–cell interactions},}\ }\href@noop {} {\bibfield  {journal}
  {\bibinfo  {journal} {Journal of the American Chemical Society}\ }\textbf
  {\bibinfo {volume} {146}},\ \bibinfo {pages} {17539--17558} (\bibinfo {year}
  {2024})}\BibitemShut {NoStop}%
\bibitem [{\citenamefont {Wilhelm}, \citenamefont {Reinheimer},\ and\
  \citenamefont {Ortseifer}(1999)}]{Wilhelm1999}%
  \BibitemOpen
  \bibfield  {author} {\bibinfo {author} {\bibfnamefont {M.}~\bibnamefont
  {Wilhelm}}, \bibinfo {author} {\bibfnamefont {P.}~\bibnamefont
  {Reinheimer}},\ and\ \bibinfo {author} {\bibfnamefont {M.}~\bibnamefont
  {Ortseifer}},\ }\bibfield  {title} {\enquote {\bibinfo {title} {High
  sensitivity fourier-transform rheology},}\ }\href
  {https://doi.org/10.1007/s003970050185} {\bibfield  {journal} {\bibinfo
  {journal} {Rheologica Acta}\ }\textbf {\bibinfo {volume} {38}},\ \bibinfo
  {pages} {349--356} (\bibinfo {year} {1999})}\BibitemShut {NoStop}%
\bibitem [{\citenamefont {Wilhelm}\ \emph {et~al.}(2000)\citenamefont
  {Wilhelm}, \citenamefont {Reinheimer}, \citenamefont {Ortseifer},
  \citenamefont {Neidhöfer},\ and\ \citenamefont {Spiess}}]{Wilhelm2000}%
  \BibitemOpen
  \bibfield  {author} {\bibinfo {author} {\bibfnamefont {M.}~\bibnamefont
  {Wilhelm}}, \bibinfo {author} {\bibfnamefont {P.}~\bibnamefont {Reinheimer}},
  \bibinfo {author} {\bibfnamefont {M.}~\bibnamefont {Ortseifer}}, \bibinfo
  {author} {\bibfnamefont {T.}~\bibnamefont {Neidhöfer}},\ and\ \bibinfo
  {author} {\bibfnamefont {H.-W.}\ \bibnamefont {Spiess}},\ }\bibfield  {title}
  {\enquote {\bibinfo {title} {The crossover between linear and non-linear
  mechanical behaviour in polymer solutions as detected by fourier-transform
  rheology},}\ }\href {https://doi.org/10.1007/s003970000084} {\bibfield
  {journal} {\bibinfo  {journal} {Rheologica Acta}\ }\textbf {\bibinfo {volume}
  {39}},\ \bibinfo {pages} {241--246} (\bibinfo {year} {2000})}\BibitemShut
  {NoStop}%
\bibitem [{\citenamefont {Wilhelm}(2002)}]{Wilhelm2002}%
  \BibitemOpen
  \bibfield  {author} {\bibinfo {author} {\bibfnamefont {M.}~\bibnamefont
  {Wilhelm}},\ }\bibfield  {title} {\enquote {\bibinfo {title}
  {Fourier-transform rheology},}\ }\href
  {https://doi.org/10.1002/1439-2054(20020201)287:2<83::aid-mame83>3.0.co;2-b}
  {\bibfield  {journal} {\bibinfo  {journal} {Macromolecular Materials and
  Engineering}\ }\textbf {\bibinfo {volume} {287}},\ \bibinfo {pages} {83--105}
  (\bibinfo {year} {2002})}\BibitemShut {NoStop}%
\bibitem [{\citenamefont {Klein}\ \emph {et~al.}(2007)\citenamefont {Klein},
  \citenamefont {Spiess}, \citenamefont {Calin}, \citenamefont {Balan},\ and\
  \citenamefont {Wilhelm}}]{Klein2007}%
  \BibitemOpen
  \bibfield  {author} {\bibinfo {author} {\bibfnamefont {C.~O.}\ \bibnamefont
  {Klein}}, \bibinfo {author} {\bibfnamefont {H.~W.}\ \bibnamefont {Spiess}},
  \bibinfo {author} {\bibfnamefont {A.}~\bibnamefont {Calin}}, \bibinfo
  {author} {\bibfnamefont {C.}~\bibnamefont {Balan}},\ and\ \bibinfo {author}
  {\bibfnamefont {M.}~\bibnamefont {Wilhelm}},\ }\bibfield  {title} {\enquote
  {\bibinfo {title} {Separation of the nonlinear oscillatory response into a
  superposition of linear, strain hardening, strain softening, and wall slip
  response},}\ }\href {https://doi.org/10.1021/ma062441u} {\bibfield  {journal}
  {\bibinfo  {journal} {Macromolecules}\ }\textbf {\bibinfo {volume} {40}},\
  \bibinfo {pages} {4250--4259} (\bibinfo {year} {2007})}\BibitemShut {NoStop}%
\bibitem [{\citenamefont {Ewoldt}, \citenamefont {Hosoi},\ and\ \citenamefont
  {McKinley}(2008)}]{Ewoldt2008}%
  \BibitemOpen
  \bibfield  {author} {\bibinfo {author} {\bibfnamefont {R.~H.}\ \bibnamefont
  {Ewoldt}}, \bibinfo {author} {\bibfnamefont {A.~E.}\ \bibnamefont {Hosoi}},\
  and\ \bibinfo {author} {\bibfnamefont {G.~H.}\ \bibnamefont {McKinley}},\
  }\bibfield  {title} {\enquote {\bibinfo {title} {New measures for
  characterizing nonlinear viscoelasticity in large amplitude oscillatory
  shear},}\ }\href {https://doi.org/10.1122/1.2970095} {\bibfield  {journal}
  {\bibinfo  {journal} {Journal of Rheology}\ }\textbf {\bibinfo {volume}
  {52}},\ \bibinfo {pages} {1427--1458} (\bibinfo {year} {2008})}\BibitemShut
  {NoStop}%
\bibitem [{\citenamefont {Ewoldt}\ \emph {et~al.}(2009)\citenamefont {Ewoldt},
  \citenamefont {Winter}, \citenamefont {Maxey},\ and\ \citenamefont
  {McKinley}}]{Ewoldt2009}%
  \BibitemOpen
  \bibfield  {author} {\bibinfo {author} {\bibfnamefont {R.~H.}\ \bibnamefont
  {Ewoldt}}, \bibinfo {author} {\bibfnamefont {P.}~\bibnamefont {Winter}},
  \bibinfo {author} {\bibfnamefont {J.}~\bibnamefont {Maxey}},\ and\ \bibinfo
  {author} {\bibfnamefont {G.~H.}\ \bibnamefont {McKinley}},\ }\bibfield
  {title} {\enquote {\bibinfo {title} {Large amplitude oscillatory shear of
  pseudoplastic and elastoviscoplastic materials},}\ }\href
  {https://doi.org/10.1007/s00397-009-0403-7} {\bibfield  {journal} {\bibinfo
  {journal} {Rheologica Acta}\ }\textbf {\bibinfo {volume} {49}},\ \bibinfo
  {pages} {191--212} (\bibinfo {year} {2009})}\BibitemShut {NoStop}%
\bibitem [{\citenamefont {Cho}\ \emph {et~al.}(2005)\citenamefont {Cho},
  \citenamefont {Hyun}, \citenamefont {Ahn},\ and\ \citenamefont
  {Lee}}]{Cho2005}%
  \BibitemOpen
  \bibfield  {author} {\bibinfo {author} {\bibfnamefont {K.~S.}\ \bibnamefont
  {Cho}}, \bibinfo {author} {\bibfnamefont {K.}~\bibnamefont {Hyun}}, \bibinfo
  {author} {\bibfnamefont {K.~H.}\ \bibnamefont {Ahn}},\ and\ \bibinfo {author}
  {\bibfnamefont {S.~J.}\ \bibnamefont {Lee}},\ }\bibfield  {title} {\enquote
  {\bibinfo {title} {A geometrical interpretation of large amplitude
  oscillatory shear response},}\ }\href {https://doi.org/10.1122/1.1895801}
  {\bibfield  {journal} {\bibinfo  {journal} {Journal of Rheology}\ }\textbf
  {\bibinfo {volume} {49}},\ \bibinfo {pages} {747--758} (\bibinfo {year}
  {2005})}\BibitemShut {NoStop}%
\bibitem [{\citenamefont {Dimitriou}, \citenamefont {Ewoldt},\ and\
  \citenamefont {McKinley}(2013)}]{Dimitriou2013}%
  \BibitemOpen
  \bibfield  {author} {\bibinfo {author} {\bibfnamefont {C.~J.}\ \bibnamefont
  {Dimitriou}}, \bibinfo {author} {\bibfnamefont {R.~H.}\ \bibnamefont
  {Ewoldt}},\ and\ \bibinfo {author} {\bibfnamefont {G.~H.}\ \bibnamefont
  {McKinley}},\ }\bibfield  {title} {\enquote {\bibinfo {title} {Describing and
  prescribing the constitutive response of yield stress fluids using large
  amplitude oscillatory shear stress (laostress)},}\ }\href
  {https://doi.org/10.1122/1.4754023} {\bibfield  {journal} {\bibinfo
  {journal} {Journal of Rheology}\ }\textbf {\bibinfo {volume} {57}},\ \bibinfo
  {pages} {27–70} (\bibinfo {year} {2013})}\BibitemShut {NoStop}%
\bibitem [{\citenamefont {Donley}\ \emph
  {et~al.}(2019{\natexlab{a}})\citenamefont {Donley}, \citenamefont {de~Bruyn},
  \citenamefont {McKinley},\ and\ \citenamefont {Rogers}}]{Donley2019}%
  \BibitemOpen
  \bibfield  {author} {\bibinfo {author} {\bibfnamefont {G.~J.}\ \bibnamefont
  {Donley}}, \bibinfo {author} {\bibfnamefont {J.~R.}\ \bibnamefont
  {de~Bruyn}}, \bibinfo {author} {\bibfnamefont {G.~H.}\ \bibnamefont
  {McKinley}},\ and\ \bibinfo {author} {\bibfnamefont {S.~A.}\ \bibnamefont
  {Rogers}},\ }\bibfield  {title} {\enquote {\bibinfo {title} {Time-resolved
  dynamics of the yielding transition in soft materials},}\ }\href
  {https://doi.org/10.1016/j.jnnfm.2018.10.003} {\bibfield  {journal} {\bibinfo
   {journal} {Journal of Non-Newtonian Fluid Mechanics}\ }\textbf {\bibinfo
  {volume} {264}},\ \bibinfo {pages} {117--134} (\bibinfo {year}
  {2019}{\natexlab{a}})}\BibitemShut {NoStop}%
\bibitem [{\citenamefont {Rogers}(2012)}]{Rogers2012}%
  \BibitemOpen
  \bibfield  {author} {\bibinfo {author} {\bibfnamefont {S.~A.}\ \bibnamefont
  {Rogers}},\ }\bibfield  {title} {\enquote {\bibinfo {title} {A sequence of
  physical processes determined and quantified in laos: An instantaneous local
  2d/3d approach},}\ }\href {https://doi.org/10.1122/1.4726083} {\bibfield
  {journal} {\bibinfo  {journal} {Journal of Rheology}\ }\textbf {\bibinfo
  {volume} {56}},\ \bibinfo {pages} {1129--1151} (\bibinfo {year}
  {2012})}\BibitemShut {NoStop}%
\bibitem [{\citenamefont {Rogers}\ \emph {et~al.}(2011)\citenamefont {Rogers},
  \citenamefont {Erwin}, \citenamefont {Vlassopoulos},\ and\ \citenamefont
  {Cloitre}}]{Rogers2011}%
  \BibitemOpen
  \bibfield  {author} {\bibinfo {author} {\bibfnamefont {S.~A.}\ \bibnamefont
  {Rogers}}, \bibinfo {author} {\bibfnamefont {B.~M.}\ \bibnamefont {Erwin}},
  \bibinfo {author} {\bibfnamefont {D.}~\bibnamefont {Vlassopoulos}},\ and\
  \bibinfo {author} {\bibfnamefont {M.}~\bibnamefont {Cloitre}},\ }\bibfield
  {title} {\enquote {\bibinfo {title} {A sequence of physical processes
  determined and quantified in laos: Application to a yield stress fluid},}\
  }\href {https://doi.org/10.1122/1.3544591} {\bibfield  {journal} {\bibinfo
  {journal} {Journal of Rheology}\ }\textbf {\bibinfo {volume} {55}},\ \bibinfo
  {pages} {435--458} (\bibinfo {year} {2011})}\BibitemShut {NoStop}%
\bibitem [{\citenamefont {Donley}, \citenamefont {Bantawa},\ and\ \citenamefont
  {Del~Gado}(2022)}]{Donley2022}%
  \BibitemOpen
  \bibfield  {author} {\bibinfo {author} {\bibfnamefont {G.~J.}\ \bibnamefont
  {Donley}}, \bibinfo {author} {\bibfnamefont {M.}~\bibnamefont {Bantawa}},\
  and\ \bibinfo {author} {\bibfnamefont {E.}~\bibnamefont {Del~Gado}},\
  }\bibfield  {title} {\enquote {\bibinfo {title} {Time-resolved
  microstructural changes in large amplitude oscillatory shear of model single
  and double component soft gels},}\ }\href {https://doi.org/10.1122/8.0000486}
  {\bibfield  {journal} {\bibinfo  {journal} {Journal of Rheology}\ }\textbf
  {\bibinfo {volume} {66}},\ \bibinfo {pages} {1287–1304} (\bibinfo {year}
  {2022})}\BibitemShut {NoStop}%
\bibitem [{\citenamefont {Rogers}\ and\ \citenamefont
  {Lettinga}(2012)}]{Rogers2012a}%
  \BibitemOpen
  \bibfield  {author} {\bibinfo {author} {\bibfnamefont {S.~A.}\ \bibnamefont
  {Rogers}}\ and\ \bibinfo {author} {\bibfnamefont {M.~P.}\ \bibnamefont
  {Lettinga}},\ }\bibfield  {title} {\enquote {\bibinfo {title} {A sequence of
  physical processes determined and quantified in large-amplitude oscillatory
  shear (laos): Application to theoretical nonlinear models},}\ }\href
  {https://doi.org/10.1122/1.3662962} {\bibfield  {journal} {\bibinfo
  {journal} {Journal of Rheology}\ }\textbf {\bibinfo {volume} {56}},\ \bibinfo
  {pages} {1--25} (\bibinfo {year} {2012})}\BibitemShut {NoStop}%
\bibitem [{\citenamefont {Kamani}\ \emph {et~al.}(2023)\citenamefont {Kamani},
  \citenamefont {Donley}, \citenamefont {Rao}, \citenamefont {Grillet},
  \citenamefont {Roberts}, \citenamefont {Shetty},\ and\ \citenamefont
  {Rogers}}]{Kamani2023}%
  \BibitemOpen
  \bibfield  {author} {\bibinfo {author} {\bibfnamefont {K.~M.}\ \bibnamefont
  {Kamani}}, \bibinfo {author} {\bibfnamefont {G.~J.}\ \bibnamefont {Donley}},
  \bibinfo {author} {\bibfnamefont {R.}~\bibnamefont {Rao}}, \bibinfo {author}
  {\bibfnamefont {A.~M.}\ \bibnamefont {Grillet}}, \bibinfo {author}
  {\bibfnamefont {C.}~\bibnamefont {Roberts}}, \bibinfo {author} {\bibfnamefont
  {A.}~\bibnamefont {Shetty}},\ and\ \bibinfo {author} {\bibfnamefont {S.~A.}\
  \bibnamefont {Rogers}},\ }\bibfield  {title} {\enquote {\bibinfo {title}
  {Understanding the transient large amplitude oscillatory shear behavior of
  yield stress fluids},}\ }\href {https://doi.org/10.1122/8.0000583} {\bibfield
   {journal} {\bibinfo  {journal} {Journal of Rheology}\ }\textbf {\bibinfo
  {volume} {67}},\ \bibinfo {pages} {331–352} (\bibinfo {year}
  {2023})}\BibitemShut {NoStop}%
\bibitem [{\citenamefont {Armstrong}\ \emph {et~al.}(2020)\citenamefont
  {Armstrong}, \citenamefont {Helton}, \citenamefont {Donley}, \citenamefont
  {Rogers},\ and\ \citenamefont {Horner}}]{Armstrong2020}%
  \BibitemOpen
  \bibfield  {author} {\bibinfo {author} {\bibfnamefont {M.}~\bibnamefont
  {Armstrong}}, \bibinfo {author} {\bibfnamefont {T.}~\bibnamefont {Helton}},
  \bibinfo {author} {\bibfnamefont {G.}~\bibnamefont {Donley}}, \bibinfo
  {author} {\bibfnamefont {S.}~\bibnamefont {Rogers}},\ and\ \bibinfo {author}
  {\bibfnamefont {J.}~\bibnamefont {Horner}},\ }\bibfield  {title} {\enquote
  {\bibinfo {title} {A small-scale study of nonlinear blood rheology shows
  rapid transient transitions},}\ }\href
  {https://doi.org/10.1007/s00397-020-01230-8} {\bibfield  {journal} {\bibinfo
  {journal} {Rheologica Acta}\ }\textbf {\bibinfo {volume} {59}},\ \bibinfo
  {pages} {687–705} (\bibinfo {year} {2020})}\BibitemShut {NoStop}%
\bibitem [{\citenamefont {Choi}, \citenamefont {Nettesheim},\ and\
  \citenamefont {Rogers}(2019)}]{Choi2019}%
  \BibitemOpen
  \bibfield  {author} {\bibinfo {author} {\bibfnamefont {J.}~\bibnamefont
  {Choi}}, \bibinfo {author} {\bibfnamefont {F.}~\bibnamefont {Nettesheim}},\
  and\ \bibinfo {author} {\bibfnamefont {S.~A.}\ \bibnamefont {Rogers}},\
  }\bibfield  {title} {\enquote {\bibinfo {title} {The unification of disparate
  rheological measures in oscillatory shearing},}\ }\href
  {https://doi.org/10.1063/1.5106378} {\bibfield  {journal} {\bibinfo
  {journal} {Physics of Fluids}\ }\textbf {\bibinfo {volume} {31}} (\bibinfo
  {year} {2019}),\ 10.1063/1.5106378}\BibitemShut {NoStop}%
\bibitem [{\citenamefont {Donley}\ \emph
  {et~al.}(2019{\natexlab{b}})\citenamefont {Donley}, \citenamefont {Hyde},
  \citenamefont {Rogers},\ and\ \citenamefont {Nettesheim}}]{Donley2019_2}%
  \BibitemOpen
  \bibfield  {author} {\bibinfo {author} {\bibfnamefont {G.~J.}\ \bibnamefont
  {Donley}}, \bibinfo {author} {\bibfnamefont {W.~W.}\ \bibnamefont {Hyde}},
  \bibinfo {author} {\bibfnamefont {S.~A.}\ \bibnamefont {Rogers}},\ and\
  \bibinfo {author} {\bibfnamefont {F.}~\bibnamefont {Nettesheim}},\ }\bibfield
   {title} {\enquote {\bibinfo {title} {Yielding and recovery of conductive
  pastes for screen printing},}\ }\href
  {https://doi.org/10.1007/s00397-019-01148-w} {\bibfield  {journal} {\bibinfo
  {journal} {Rheologica Acta}\ }\textbf {\bibinfo {volume} {58}},\ \bibinfo
  {pages} {361–382} (\bibinfo {year} {2019}{\natexlab{b}})}\BibitemShut
  {NoStop}%
\bibitem [{\citenamefont {Korculanin}\ \emph {et~al.}(2021)\citenamefont
  {Korculanin}, \citenamefont {Westermeier}, \citenamefont {Hirsemann},
  \citenamefont {Struth}, \citenamefont {Hermida-Merino}, \citenamefont
  {Wagner}, \citenamefont {Donley}, \citenamefont {Rogers},\ and\ \citenamefont
  {Lettinga}}]{Korculanin2021}%
  \BibitemOpen
  \bibfield  {author} {\bibinfo {author} {\bibfnamefont {O.}~\bibnamefont
  {Korculanin}}, \bibinfo {author} {\bibfnamefont {F.}~\bibnamefont
  {Westermeier}}, \bibinfo {author} {\bibfnamefont {H.}~\bibnamefont
  {Hirsemann}}, \bibinfo {author} {\bibfnamefont {B.}~\bibnamefont {Struth}},
  \bibinfo {author} {\bibfnamefont {D.}~\bibnamefont {Hermida-Merino}},
  \bibinfo {author} {\bibfnamefont {U.~H.}\ \bibnamefont {Wagner}}, \bibinfo
  {author} {\bibfnamefont {G.~J.}\ \bibnamefont {Donley}}, \bibinfo {author}
  {\bibfnamefont {S.~A.}\ \bibnamefont {Rogers}},\ and\ \bibinfo {author}
  {\bibfnamefont {M.~P.}\ \bibnamefont {Lettinga}},\ }\bibfield  {title}
  {\enquote {\bibinfo {title} {Anomalous dynamic response of nematic platelets
  studied by spatially resolved rheo-small angle x-ray scattering in the 1–2
  plane},}\ }\href {https://doi.org/10.1063/5.0069458} {\bibfield  {journal}
  {\bibinfo  {journal} {Physics of Fluids}\ }\textbf {\bibinfo {volume} {33}}
  (\bibinfo {year} {2021}),\ 10.1063/5.0069458}\BibitemShut {NoStop}%
\bibitem [{\citenamefont {Lee}\ and\ \citenamefont {Rogers}(2017)}]{Lee2017}%
  \BibitemOpen
  \bibfield  {author} {\bibinfo {author} {\bibfnamefont {C.-W.}\ \bibnamefont
  {Lee}}\ and\ \bibinfo {author} {\bibfnamefont {S.~A.}\ \bibnamefont
  {Rogers}},\ }\bibfield  {title} {\enquote {\bibinfo {title} {A sequence of
  physical processes quantified in laos by continuous local measures},}\ }\href
  {https://doi.org/10.1007/s13367-017-0027-x} {\bibfield  {journal} {\bibinfo
  {journal} {Korea-Australia Rheology Journal}\ }\textbf {\bibinfo {volume}
  {29}},\ \bibinfo {pages} {269--279} (\bibinfo {year} {2017})}\BibitemShut
  {NoStop}%
\bibitem [{\citenamefont {Dubbin}\ \emph {et~al.}(2020)\citenamefont {Dubbin},
  \citenamefont {Robertson}, \citenamefont {Hinckley}, \citenamefont
  {Alvarado}, \citenamefont {Gilmore}, \citenamefont {Hynes}, \citenamefont
  {Wheeler},\ and\ \citenamefont {Moya}}]{Dubbin2020}%
  \BibitemOpen
  \bibfield  {author} {\bibinfo {author} {\bibfnamefont {K.}~\bibnamefont
  {Dubbin}}, \bibinfo {author} {\bibfnamefont {C.}~\bibnamefont {Robertson}},
  \bibinfo {author} {\bibfnamefont {A.}~\bibnamefont {Hinckley}}, \bibinfo
  {author} {\bibfnamefont {J.~A.}\ \bibnamefont {Alvarado}}, \bibinfo {author}
  {\bibfnamefont {S.~F.}\ \bibnamefont {Gilmore}}, \bibinfo {author}
  {\bibfnamefont {W.~F.}\ \bibnamefont {Hynes}}, \bibinfo {author}
  {\bibfnamefont {E.~K.}\ \bibnamefont {Wheeler}},\ and\ \bibinfo {author}
  {\bibfnamefont {M.~L.}\ \bibnamefont {Moya}},\ }\bibfield  {title} {\enquote
  {\bibinfo {title} {Macromolecular gelatin properties affect fibrin
  microarchitecture and tumor spheroid behavior in fibrin-gelatin gels},}\
  }\href {https://doi.org/10.1016/j.biomaterials.2020.120035} {\bibfield
  {journal} {\bibinfo  {journal} {Biomaterials}\ }\textbf {\bibinfo {volume}
  {250}},\ \bibinfo {pages} {120035} (\bibinfo {year} {2020})}\BibitemShut
  {NoStop}%
\bibitem [{\citenamefont {Afewerki}\ \emph {et~al.}(2018)\citenamefont
  {Afewerki}, \citenamefont {Sheikhi}, \citenamefont {Kannan}, \citenamefont
  {Ahadian},\ and\ \citenamefont {Khademhosseini}}]{Afewerki2018}%
  \BibitemOpen
  \bibfield  {author} {\bibinfo {author} {\bibfnamefont {S.}~\bibnamefont
  {Afewerki}}, \bibinfo {author} {\bibfnamefont {A.}~\bibnamefont {Sheikhi}},
  \bibinfo {author} {\bibfnamefont {S.}~\bibnamefont {Kannan}}, \bibinfo
  {author} {\bibfnamefont {S.}~\bibnamefont {Ahadian}},\ and\ \bibinfo {author}
  {\bibfnamefont {A.}~\bibnamefont {Khademhosseini}},\ }\bibfield  {title}
  {\enquote {\bibinfo {title} {Gelatin‐polysaccharide composite scaffolds for
  3d cell culture and tissue engineering: Towards natural therapeutics},}\
  }\href {https://doi.org/10.1002/btm2.10124} {\bibfield  {journal} {\bibinfo
  {journal} {Bioengineering \& Translational Medicine}\ }\textbf {\bibinfo
  {volume} {4}},\ \bibinfo {pages} {96–115} (\bibinfo {year}
  {2018})}\BibitemShut {NoStop}%
\bibitem [{\citenamefont {Gardel}\ \emph {et~al.}(2004)\citenamefont {Gardel},
  \citenamefont {Shin}, \citenamefont {MacKintosh}, \citenamefont {Mahadevan},
  \citenamefont {Matsudaira},\ and\ \citenamefont {Weitz}}]{Gardel2004}%
  \BibitemOpen
  \bibfield  {author} {\bibinfo {author} {\bibfnamefont {M.~L.}\ \bibnamefont
  {Gardel}}, \bibinfo {author} {\bibfnamefont {J.~H.}\ \bibnamefont {Shin}},
  \bibinfo {author} {\bibfnamefont {F.~C.}\ \bibnamefont {MacKintosh}},
  \bibinfo {author} {\bibfnamefont {L.}~\bibnamefont {Mahadevan}}, \bibinfo
  {author} {\bibfnamefont {P.}~\bibnamefont {Matsudaira}},\ and\ \bibinfo
  {author} {\bibfnamefont {D.~A.}\ \bibnamefont {Weitz}},\ }\bibfield  {title}
  {\enquote {\bibinfo {title} {Elastic behavior of cross-linked and bundled
  actin networks},}\ }\href {https://doi.org/10.1126/science.1095087}
  {\bibfield  {journal} {\bibinfo  {journal} {Science}\ }\textbf {\bibinfo
  {volume} {304}},\ \bibinfo {pages} {1301--1305} (\bibinfo {year}
  {2004})}\BibitemShut {NoStop}%
\bibitem [{\citenamefont {Bischoff}, \citenamefont {Arruda},\ and\
  \citenamefont {Grosh}(2004)}]{Bischoff2004}%
  \BibitemOpen
  \bibfield  {author} {\bibinfo {author} {\bibfnamefont {J.~E.}\ \bibnamefont
  {Bischoff}}, \bibinfo {author} {\bibfnamefont {E.~M.}\ \bibnamefont
  {Arruda}},\ and\ \bibinfo {author} {\bibfnamefont {K.}~\bibnamefont
  {Grosh}},\ }\bibfield  {title} {\enquote {\bibinfo {title} {A rheological
  network model for the continuum anisotropic and viscoelastic behavior of soft
  tissue},}\ }\href {https://doi.org/10.1007/s10237-004-0049-4} {\bibfield
  {journal} {\bibinfo  {journal} {Biomechanics and Modeling in Mechanobiology}\
  }\textbf {\bibinfo {volume} {3}},\ \bibinfo {pages} {56--65} (\bibinfo {year}
  {2004})}\BibitemShut {NoStop}%
\bibitem [{\citenamefont {Janmey}, \citenamefont {Winer},\ and\ \citenamefont
  {Weisel}(2008)}]{Janmey2008}%
  \BibitemOpen
  \bibfield  {author} {\bibinfo {author} {\bibfnamefont {P.~A.}\ \bibnamefont
  {Janmey}}, \bibinfo {author} {\bibfnamefont {J.~P.}\ \bibnamefont {Winer}},\
  and\ \bibinfo {author} {\bibfnamefont {J.~W.}\ \bibnamefont {Weisel}},\
  }\bibfield  {title} {\enquote {\bibinfo {title} {Fibrin gels and their
  clinical and bioengineering applications},}\ }\href
  {https://doi.org/10.1098/rsif.2008.0327} {\bibfield  {journal} {\bibinfo
  {journal} {Journal of The Royal Society Interface}\ }\textbf {\bibinfo
  {volume} {6}},\ \bibinfo {pages} {1--10} (\bibinfo {year}
  {2008})}\BibitemShut {NoStop}%
\bibitem [{\citenamefont {Kang}\ \emph {et~al.}(2009)\citenamefont {Kang},
  \citenamefont {Wen}, \citenamefont {Janmey}, \citenamefont {Tang},
  \citenamefont {Conti},\ and\ \citenamefont {MacKintosh}}]{Kang2009}%
  \BibitemOpen
  \bibfield  {author} {\bibinfo {author} {\bibfnamefont {H.}~\bibnamefont
  {Kang}}, \bibinfo {author} {\bibfnamefont {Q.}~\bibnamefont {Wen}}, \bibinfo
  {author} {\bibfnamefont {P.~A.}\ \bibnamefont {Janmey}}, \bibinfo {author}
  {\bibfnamefont {J.~X.}\ \bibnamefont {Tang}}, \bibinfo {author}
  {\bibfnamefont {E.}~\bibnamefont {Conti}},\ and\ \bibinfo {author}
  {\bibfnamefont {F.~C.}\ \bibnamefont {MacKintosh}},\ }\bibfield  {title}
  {\enquote {\bibinfo {title} {Nonlinear elasticity of stiff filament networks:
  Strain stiffening, negative normal stress, and filament alignment in fibrin
  gels},}\ }\href {https://doi.org/10.1021/jp807749f} {\bibfield  {journal}
  {\bibinfo  {journal} {The Journal of Physical Chemistry B}\ }\textbf
  {\bibinfo {volume} {113}},\ \bibinfo {pages} {3799--3805} (\bibinfo {year}
  {2009})}\BibitemShut {NoStop}%
\bibitem [{\citenamefont {Janmey}, \citenamefont {Amis},\ and\ \citenamefont
  {Ferry}(1983)}]{janmey_jor_83}%
  \BibitemOpen
  \bibfield  {author} {\bibinfo {author} {\bibfnamefont {P.~A.}\ \bibnamefont
  {Janmey}}, \bibinfo {author} {\bibfnamefont {E.~J.}\ \bibnamefont {Amis}},\
  and\ \bibinfo {author} {\bibfnamefont {J.~D.}\ \bibnamefont {Ferry}},\
  }\bibfield  {title} {\enquote {\bibinfo {title} {{Rheology of Fibrin Clots.
  VI. Stress Relaxation, Creep, and Differential Dynamic Modulus of Fine Clots
  in Large Shearing Deformations}},}\ }\href {https://doi.org/10.1122/1.549722}
  {\bibfield  {journal} {\bibinfo  {journal} {Journal of Rheology}\ }\textbf
  {\bibinfo {volume} {27}},\ \bibinfo {pages} {135--153} (\bibinfo {year}
  {1983})},\ \Eprint
  {https://arxiv.org/abs/https://pubs.aip.org/sor/jor/article-pdf/27/2/135/12690473/135\_1\_online.pdf}
  {https://pubs.aip.org/sor/jor/article-pdf/27/2/135/12690473/135\_1\_online.pdf}
  \BibitemShut {NoStop}%
\bibitem [{\citenamefont {Shah}\ and\ \citenamefont
  {Janmey}(1997)}]{shah_strain_1997}%
  \BibitemOpen
  \bibfield  {author} {\bibinfo {author} {\bibfnamefont {J.~V.}\ \bibnamefont
  {Shah}}\ and\ \bibinfo {author} {\bibfnamefont {P.~A.}\ \bibnamefont
  {Janmey}},\ }\bibfield  {title} {\enquote {\bibinfo {title} {Strain hardening
  of fibrin gels and plasma clots},}\ }\href
  {https://doi.org/10.1007/BF00366667} {\bibfield  {journal} {\bibinfo
  {journal} {Rheologica Acta}\ }\textbf {\bibinfo {volume} {36}},\ \bibinfo
  {pages} {262--268} (\bibinfo {year} {1997})}\BibitemShut {NoStop}%
\bibitem [{\citenamefont {Piechocka}\ \emph
  {et~al.}(2010{\natexlab{a}})\citenamefont {Piechocka}, \citenamefont
  {Bacabac}, \citenamefont {Potters}, \citenamefont {MacKintosh},\ and\
  \citenamefont {Koenderink}}]{piechocka_structural_2010}%
  \BibitemOpen
  \bibfield  {author} {\bibinfo {author} {\bibfnamefont {I.~K.}\ \bibnamefont
  {Piechocka}}, \bibinfo {author} {\bibfnamefont {R.~G.}\ \bibnamefont
  {Bacabac}}, \bibinfo {author} {\bibfnamefont {M.}~\bibnamefont {Potters}},
  \bibinfo {author} {\bibfnamefont {F.~C.}\ \bibnamefont {MacKintosh}},\ and\
  \bibinfo {author} {\bibfnamefont {G.~H.}\ \bibnamefont {Koenderink}},\
  }\bibfield  {title} {\enquote {\bibinfo {title} {Structural {Hierarchy}
  {Governs} {Fibrin} {Gel} {Mechanics}},}\ }\href
  {https://doi.org/10.1016/j.bpj.2010.01.040} {\bibfield  {journal} {\bibinfo
  {journal} {Biophysical Journal}\ }\textbf {\bibinfo {volume} {98}},\ \bibinfo
  {pages} {2281--2289} (\bibinfo {year} {2010}{\natexlab{a}})},\ \bibinfo
  {note} {publisher: Elsevier}\BibitemShut {NoStop}%
\bibitem [{\citenamefont {Donley}(2021)}]{Donley2021}%
  \BibitemOpen
  \bibfield  {author} {\bibinfo {author} {\bibfnamefont {G.~J.}\ \bibnamefont
  {Donley}},\ }\emph {\bibinfo {title} {The yielding transition in soft
  materials}},\ \href@noop {} {Ph.D. thesis},\ \bibinfo  {school} {The
  University of Illinois at Urbana-Champaign} (\bibinfo {year}
  {2021})\BibitemShut {NoStop}%
\bibitem [{\citenamefont {Piechocka}\ \emph
  {et~al.}(2010{\natexlab{b}})\citenamefont {Piechocka}, \citenamefont
  {Bacabac}, \citenamefont {Potters}, \citenamefont {MacKintosh},\ and\
  \citenamefont {Koenderink}}]{PIECHOCKA20102281}%
  \BibitemOpen
  \bibfield  {author} {\bibinfo {author} {\bibfnamefont {I.~K.}\ \bibnamefont
  {Piechocka}}, \bibinfo {author} {\bibfnamefont {R.~G.}\ \bibnamefont
  {Bacabac}}, \bibinfo {author} {\bibfnamefont {M.}~\bibnamefont {Potters}},
  \bibinfo {author} {\bibfnamefont {F.~C.}\ \bibnamefont {MacKintosh}},\ and\
  \bibinfo {author} {\bibfnamefont {G.~H.}\ \bibnamefont {Koenderink}},\
  }\bibfield  {title} {\enquote {\bibinfo {title} {Structural hierarchy governs
  fibrin gel mechanics},}\ }\href
  {https://doi.org/https://doi.org/10.1016/j.bpj.2010.01.040} {\bibfield
  {journal} {\bibinfo  {journal} {Biophysical Journal}\ }\textbf {\bibinfo
  {volume} {98}},\ \bibinfo {pages} {2281--2289} (\bibinfo {year}
  {2010}{\natexlab{b}})}\BibitemShut {NoStop}%
\end{thebibliography}%

\end{document}